\documentclass[12pt,preprint]{aastex}

\def\ga{\mathrel{\raise.3ex\hbox{$>$\kern-.75em\lower1ex\hbox{$\sim$}}}}
\def\la{\mathrel{\raise.3ex\hbox{$<$\kern-.75em\lower1ex\hbox{$\sim$}}}}

\newcommand\beq{\begin{equation}}
\newcommand\eeq{\end{equation}}
\newcommand\beqar{\begin{eqnarray}}
\newcommand\eeqar{\end{eqnarray}}
\usepackage{epsfig}

\begin{document}

\vskip 0.2in
\title{The Fine-structure Constant as a Probe of  Chemical Evolution and AGB Nucleosynthesis
in Damped Lyman-$\alpha$ Systems}
\author{Timothy P. Ashenfelter, Grant J. Mathews}
\affil{Department of Physics, Center for Astrophysics (CANDU) and Joint Institute for Nuclear Astrophysics (JINA), \\
University of Notre Dame, Notre Dame, IN, 46556}

\author{Keith A. Olive}
\affil{William I. Fine Theoretical Physics Institute, \\
University of Minnesota, Minneapolis, MN 55455, USA}

\begin{abstract}
\vskip-4.6in
\begin{flushright}
UMN-TH-2302/04 \\
TPI-MINN-04/13 \\
astro-ph/0404257 \\
April 2004
\end{flushright}
\vskip+3.6in
Evidence from a large sample of quasar absorption-line spectra in damped
Lyman-$\alpha$ systems has suggested a 
possible time variation of the fine structure constant $\alpha$. 
The most statistically significant portion of this sample involves the comparison of 
Mg and Fe wavelength shifts using the many-multiplet (MM) method. 
However, the sensitivity of this method to the abundance of heavy isotopes, 
especially Mg, is enough to imitate an apparent variation in $\alpha$ in the
redshift range $0.5 < z < 1.8$.  
We implement recent yields of  intermediate mass (IM)  stars into a chemical evolution model 
and show that the ensuing isotope distribution of Mg can account 
for the observed variation in $\alpha$
provided the early IMF was particularly rich in intermediate mass stars
(or the  heavy Mg isotope yields from AGB stars are even higher than in
 present-day models).  As such, these observations of quasar absorption spectra
can be used to probe the nucleosynthetic history of
low-metallicity damped Lyman-$\alpha$ systems in the redshift range
$0.5 < z < 1.8$. 
This analysis,  in conjunction with other abundance measurements of low-metallicity systems, 
reinforces the mounting evidence that star formation at low
metallicities may have been strongly influenced by a population of         IM         stars.
Such        IM         stars have a significant influence on other abundances, 
particularly nitrogen. We constrain our models with independent measurements of 
N, Si, and Fe in damped Lyman-alpha systems as well as C/O in low-metallicity stars.
In this way, we obtain consistent model
parameters for this chemical-evolution interpretation of the 
MM method results.
\end{abstract}



\section{Introduction}
The origin and dynamics of the fundamental constants of Nature is
one of the deepest questions in physics.
 One of the most widely held tenets in physics is that the laws of 
nature are universal, constant, and favor symmetries. 
Nevertheless,  in various  unified theories (including string theory), 
gauge and Yukawa couplings 
often appear as dynamical variables which are only ``fixed" 
when a related scalar field (such as a dilaton in string theory) picks up a vacuum
expectation value. While one may naturally expect that couplings become constant 
at or near the unification scale, only experimental evidence can constrain 
the degree to which these constants vary at late times.

In this context there has been considerable excitement in recent years
over the prospect that a time variation in the 
dimensionless fine structure constant, $\alpha$, 
may have been observed 
 (Webb et al.~1999, Murphy et al.~2001a,b, Murphy et al.~2003a).  This evidence is 
 based upon an application of the ``many multiplet" (MM)
 method to quasar absorption lines in damped Lyman-$\alpha$ systems (DLAs).
Attempts at constraining or measuring time variations of $\alpha$ in quasar 
absorption-line spectra have 
a long history going back to work by Bahcall \& Salpeter (1965) 
using O$III$ and Ne$III$ emission lines. 
This method was reexamined recently  by Bahcall, Steinhardt, \& Schlegel (2004). 
Other recent attempts include measurements of absorption line spectra in alkali-like atoms 
(Potekhin  \& Varshalovich 1994, Varshalovich \& Potekhin 1994, Murphy et al.~2001c,
Fiorenzano, Vladilo, \& Bonifacio 2003). While many observations have led to interesting 
limits on the temporal variation of $\alpha$ (for a recent review, see Uzan 2003), 
only the MM method has led to a quantitatively positive result.
Murphy et al.~(2003a) deduce
that ${\delta \alpha  /\alpha}  = (-0.54 \pm 0.12) \times 10^{-5}$
over a redshift range of $0.5 < z < 3$, where $\delta \alpha$ is defined 
as the deviation from present value. The implications of this deduced variation
in $\alpha$ at around a 5$\sigma$ significance are phenomenal, and 
several cosmological models to
explain its origin have been proposed
(see e.g., Beckenstein 1982, Sandvik, Barrow, \& Magueijo 2002, Olive \& Pospelov 2002,
Wetterich 2003, Anchordoqui \& Goldberg 2003, Copeland, Nunes, \& Pospelov 2004,
Lee, Lee, \& Ng 2003; Byrne \& Kolda 2004). The caveat of implementing such a precise 
method, however, is its possible sensitivity to unknown systematic errors. 

Some of the excitement concerning the evolution of the fine structure constant has been focused on finding alternative explanations of the observed line structures or other systematic errors.  Chand et al.~(2004) and Srianand et al.~(2004) probed the sensitivity of the MM method with respect to synthetic signal alterations and found that the MM method may break down in well blended, multi-cloud systems. They also applied the MM method independently and found   ${\delta \alpha  / \alpha}  = (-0.06 \pm 0.06) \times 10^{-5}$. 
Another group (Quast, Reimers, \& Levshakov
 2004) has also recently applied the MM method utilizing 
exceptionally high-resolution QSO absorption-line spectra.
 Their results are also consistent with 
a null hypothesis regarding the fine structure evolution. 
Other potential systematic errors in the MM method have been elicited by others.
They  involve the cloud velocity structure and line blending 
(Bahcall et al.~2004) or  cloud inhomogeneity and spectrographic inconsistencies (Levshakov 2003).

A number of sources of possible systematic error in this method have been well documented 
(Murphy et al.~2001b and 2003b; see also Bahcall et al.~2004). 
Here, we will focus on one of these possible systematic errors for which 
there is recent evidence of a new interpretation, namely the isotopic 
abundances of Mg assumed in the analysis. 
In this paper,  we expand on our earlier studies (Ashenfelter, Mathews \& Olive 2004)
of possible systematic effects from the chemical evolution of
magnesium isotopes within DLA quasar absorption-line systems.  
All of the results quoted above are based on the {\em assumption} that the
isotopic abundance ratios of Mg are Solar in these systems.
Based upon galactic chemical evolution studies previously available 
(Timmes, Woosley \& Weaver 1995), one could argue that the ratio of $^{25,26}$Mg/$^{24}$Mg 
is expected to decrease at low metallicity. In this case, 
if it is assumed that only 
$^{24}$Mg is present in the 
absorbers, the Murphy et al.~(2003a) result becomes significantly stronger
${\delta \alpha  / \alpha}  = (-0.86 \pm 0.10) \times 10^{-5}$ (assuming also
that only $^{28}$Si is present) and the Chand et al.~(2004)
limit becomes a detection  ${\delta \alpha  / \alpha}  = (-0.36 \pm 0.06) \times 10^{-5}$.
Hence, it is possible that the detections of time-varying $\alpha$ are even more significant
than the quoted confidence limits.
However, we show that it is also plausible that the 
$^{25,26}$Mg/$^{24}$Mg ratio was in fact sufficiently {\em higher} at low metallicity
to account for the apparent variation in $\alpha$ as seen
in the so-called ``low redshift"  ($0.5 < z < 1.8$) data.  
Thus,  the MM method of analysis
may provide important new insights into the chemical evolution
of damped Lyman-$\alpha$ quasar absorption-line systems rather than conclusive evidence
for a time-varying fine-structure constant. 

We begin the present discussion by briefly reviewing the current 
observational limits on the variations of the
fine structure constant.  In section 3, we discuss the theory and observations
of the Mg isotopes. The sensitivity of $(\delta \alpha / \alpha)$ to the Mg isotopic abundances is
explained in section 4.  In section 5, we describe
a simple chemical evolution model,
which we then utilize to address the question of the history of the Mg isotopes
and other elements
as a function of metallicity.
  Results of this study  and the sensitivity to the model are discussed
in section 6. Our summary and conclusions are given in section 7.
 
\section{Limits on the fine-structure evolution}

There are a number of important astrophysical and terrestrial constraints 
on the fine-structure constant that must be respected. 
The most primordial of the limits comes from big bang nucleosynthesis (Kolb, Perry, \& Walker 1986, 
Malaney \& Mathews 1993; Scherrer \& Spergel 1993, Campbell
\& Olive 1995, Bergstrom, Iguri, \& Rubinstein 1999, Nollett \& Lopez 2002), 
which tests for variations back to a redshift as high as $\sim$ $10^{10}$. 
However, the limit attained is rather weak  $(\delta \alpha  / \alpha) \leq 10^{-2}$. 
If one assumes that variations in the fine structure constant are accompanied by variations 
in other gauge and Yukawa couplings (Campbell \& Olive 1995, Langacker, Segre, \& Strassler 2002, Dent \& Fairbairn
 2002, Calmet \& Fritzsch 2003, Damour, Piazza \& Veneziano 2002), 
this limit can be strengthened by about two orders of magnitude (Campbell 
\& Olive 1995, Ichikawa \& Kawasaki 2002). 
There is also a slightly weaker constraint $(\delta \alpha  / \alpha) \leq$ few $\times 10^{-2}$
from the {\it WMAP}  cosmic microwave background power spectrum 
corresponding to the epoch of photon last scattering at a redshift of $z \approx 1100$
 (Rocha et al.~2003).

At smaller look-back times (lower redshifts), there are significantly stronger limits. 
Meteoritic data on the radioactive $\beta$-decay of $^{187}$Re was used 
to place an upper bound of  
${\delta \alpha / \alpha} \leq 10^{-7}$ 
(Olive et al.~2002, 2004, Fujii \& Iwamoto 2004) and is applicable 
to a redshift of $ z \leq 0.45 $. 
This limit improves by a factor of $\sim$ 25 when variations in $\alpha$ 
are assumed to be coupled  with variations in other physical constants. 
The strongest limit is based on the $^{149}$Sm resonant neutron-capture cross-section 
operating in the Oklo natural fission reactor 
(Shlyakhter 1976, Damour \& Dyson 1996, Fujii et al.~ 2000) 
about 2 billion years ago ($z \sim 0.15 $). 
The limit on $\alpha$ is  ${\delta \alpha /     \alpha} \le 10^{-7}$ 
and can be improved by 2-3 orders of magnitude 
when coupled with variations in other constants (Olive et al.~2002). 
Finally, atomic clocks provide very stringent constraints on
the present-day rate of change of $\alpha$.  
By comparing hyperfine transitions in $^{87}$Rb and $^{133}$Cs using 
atomic fountains over a period of 5 years, Marion et al.~(2003)
were able to derive the limit $\dot \alpha / \alpha < 1.5 \times 10^{-15}$ yr$^{-1}$. 
Combined with data from Hg$^+$ and H, Fischer et al.~(2003) obtain a slightly stronger 
bound $\delta \alpha / \alpha = (-0.9 \pm 4.2) \times 10^{-15}$  over a 3-4 year period.
For a comprehensive review, see Uzan (2003).

Interestingly enough, none of the limits above cover the redshift range corresponding 
to the quasar absorption-line DLAs that yield the recent 
evidence of a variation in $\alpha$.  Hence, the door remains open
for creative model building. Nevertheless, the strength of these limits, 
particularly the Oklo and $^{187}$Re bounds, may indicate that something other than 
a time-varying $\alpha$ may be responsible for the effects seen in the many-multiplet method.

\section{Mg: Theory and Observations}

\subsection{Expected trends of Mg production}

The MM method is sensitive 
to the Mg isotope abundances for data in the $0.5 < z < 1.8$ redshift range
for which the apparent variation in $\alpha$ is most pronounced  
(Murphy et al.~2003b).
Hence, it is imperative to carefully scrutinize the chemical-evolution history that one 
expects to occur within the QSO absorption systems.  
The DLAs that 
were studied to obtain the apparent variation in the 
fine-structure constant are likely to be galaxies in various stages of evolution.  
Fenner, Prochaska \& Gibson (2004) conclude that 
the sources of DLAs span a wide range of galaxy morphological types and sizes, 
from dwarf irregulars to giant ellipticals. Hence, chemical evolution models that fall 
within these broad morphologies can be used to 
explore the expected trends in the abundance distributions. 

Mg is produced in both Type I and Type II supernovae. 
In Type II supernovae, it is produced in the carbon and neon burning shells with an abundance 
somewhat less than 10 percent of the oxygen abundance produced
in massive stars (Woosley \& Weaver 1995; henceforth WW95). 
$^{25,26}$Mg are produced primarily in the outer carbon layer by 
the reactions $^{22}$Ne($\alpha$,n)$^{25}$Mg and $^{25}$Mg(n,$\gamma$)$^{26}$Mg. 
The Solar-metallicity  models of WW95  eject
Solar values with
Mg isotopes ratios reasonably close to the terrestrially observed value of $^{24}$Mg:$^{25}$Mg:$^{26}$Mg = 79:10:11  (Rosman \& Taylor 1998).
More massive stars tend to be slightly enhanced in the heavy isotopes 
(eg., the WW95 25 $M_\odot$ model gives a ratio of 65:15:20).  
Furthermore, the abundance of  $^{25,26}$Mg scales linearly with metallicity in the carbon shell. 
As a result, it would be expected that the ejecta from the first generation of 
Type-II supernovae would contribute almost no $^{25,26}$Mg.  
Some of the Solar abundance of Mg is produced in Type I supernovae but 
with the [Mg/Fe] far below Solar.  
For example, the models of Thielemann, Nomoto, \& Yokoi 
(1986) give [Mg/Fe] $\cong$ -1.2. 
Due to the absence of free neutrons, essentially no  $^{25,26}$Mg  is produced in Type I supernovae. 
Thus, in previous chemical evolution models the heavy Mg isotopes could only be efficiently produced
at late times in high metallicity Type II supernovae.

Moreover, of critical importance in the present study is that 
$^{25,26}$Mg can also be produced in intermediate-mass (IM)
 asymptotic-giant-branch (AGB) stars.
It has been recently noted (Karakas \& Lattanzio 2003; 
Siess, Livio, \& Lattanzio 2002; Forestini \& Charbonnel 1996)  
that IM stars of low metallicity can in fact be {\it efficient} producers 
of the heavy Mg isotopes during the thermal-pulsing AGB phase. 

It should be noted that the Mg yields of WW95 
near 20 M$_\odot$ are substantially lower than the predictions of other
stellar models as is described in Argast, Samland, Thielemann, 
\& Gerhard (2002). While this underestimate only affects a narrow mass range in
the Type II yields, it may have an impact on the total Mg yield in IMF weighted
chemical evolution models. If the $^{24}Mg$ production from the Type-II models
were corrected upward with respect to the WW95 yields, the expected heavy
isotopic ratio at low metallicity would decrease. $^{25,26}Mg$ scales with
the metallicity, so it would not be affected by this uncertainty. The primary
consequence of the $^{24}Mg$ underestimate in this mass range would be to slightly
enhance the impact that the Mg yields from AGB stars on the Mg isotopic ratio.
However, an enhancement of the Type-II Mg yields could be 
compensated by an appropriate adjustment to the IMF as described below in 
Section 5.  

\subsection{Mg production in AGB stars}

Heavy magnesium isotopes are synthesized via two mechanisms both of 
which are particularly robust in 2.5 - 6 $M_\odot$ stars with low metallicity. 
Such low-metallicity stars are precisely the kinds of objects which 
ought to produce the abundances observed in damped Lyman-$\alpha$ 
QSO absorption-line systems at high redshift for the reasons already described. 

One process particularly effective in low-metallicity AGB stars 
(Boothroyd, Sackmann \& Wasserburg 1995) is that of hot-bottom burning (HBB). 
During the AGB phase, stars develop an extended outer convective envelope 
and material in the hydrogen envelope
 is mixed downward to regions of high temperature at the base. 
Of particular interest is that the base of the 
envelope is more compact and at higher temperature in low-metallicity stars 
than in stars of Solar composition. This can be traced to the decreased opacity of these objects. 
Because these stars become sufficiently hot ($T > 7 \times 10^7$ K), 
proton capture processes in the Mg-Al cycle become effective. 
Proton capture on $^{24}$Mg then leads to the production of $^{25}$Mg 
(from the decay of $^{25}$Al) and to $^{26}$Al (which decays to $^{26}$Mg). 
The  relevant thermonuclear burning  reactions have a strong sensitivity 
to temperature. Hence, IM progenitor 
stars of low metallicity can contribute abundant products from high-temperature burning
to the interstellar medium. Moreover, these stars should also have a shorter 
timescale than their higher-metallicity counterparts, though this effect is 
not large as 5-6 M$_\odot$ stars have relatively short lifetimes at any
metallicity. 

A second contributing process occurs deeper in the star during thermal pulses 
of the helium-burning shell. The helium shell experiences periodic thermonuclear runaways 
when the ignition of the triple-alpha reaction occurs under electron-degenerate conditions. 
Due to electron degeneracy, the star is unable to expand and cool. 
Hence, the temperature rapidly rises until the onset of convection 
to transport the energy away. During these thermal pulses, 
$^{22}$Ne is produced by $\alpha$-capture on $^{14}$N, 
which itself is left over from the CNO cycle.
Heavy magnesium isotopes are then produced via the  
$^{22}$Ne($\alpha$,n)$^{25}$Mg and   $^{22}$Ne($\alpha,\gamma$)$^{26}$Mg reactions.  
Note that this process is important for stars with $M \approx 3$ M$_\odot$ (Karakas \& Lattanzio 2003), while HBB is more important for $M \ga 4$ M$_\odot$.

Several groups give credence to the assertion that AGB stars produce heavy Mg isotopes. 
Siess, Livio, \& Lattanzio (2002) describe the $^{25,26}$Mg production that 
occurs during the 3rd dredge-up.  
A key point is that even though seed material is less plentiful in low-metallicity stars, 
the reactions are very temperature sensitive.  
Hence, the increased temperature in the interior of low-metallicity stars 
more than compensates for the depleted seed material, 
leading to significant production of the heavy Mg isotopes. 
Moreover, seed material produced and mixed during the first 
two dredge-up episodes will also be more efficiently produced due to 
the heightened temperature.
It has even been argued that these processes may also be net 
destroyers of $^{24}$Mg (Forestini \& Charbonnel 1996, Karakas \& Lattanzio 2003) 
due to the extreme temperatures attained. 

To illustrate this phenomenon regarding the Mg evolution in AGB stars, 
Denissenkov \& Herwig (2003) modeled a typical low-metallicity ($Z=0.0001$) 
AGB star of 5 $M_\odot$ and 
found that it was capable of taking an initial ratio of $^{24}$Mg:$^{25}$Mg:$^{26}$Mg = 90:5:5 
to the extreme ratio of (13:71:16)! 
It should be noted that  a Mg isotopic composition this enriched in $^{25}$Mg has never been detected.
Nevertheless, their model establishes an upper limit on the effect of HBB 
on material dredged up to the surface.
 The main impetus of their study was to qualitatively illustrate 
that neither low seed material nor recently updated reaction rates can 
prevent the efficient processing of Mg in AGB stars.   

\subsection{Observations of Mg abundances}
The data on Mg abundances in low-metallicity stars exhibits considerable dispersion. 
This dispersion is in excess of a factor of 3 for a fixed value of [Fe/H]. 
While such dispersion could be a symptom of systematic error when data 
from several samples are combined, the observed intrinsic dispersion in abundances of
low-metallicity stars is generally interpreted (e.g. Ishimaru \& Wanajo 1999)
as evidence for effects  of  local stochastic star-formation events. 
(We note, however,  that the recent data of Cayrel et al. (2004) show considerably
less dispersion in [Mg/Fe] and other [$\alpha$/Fe] ratios in halo stars than previous
results.  Hence, the need for highly inhomogeneous halo evolution
is correspondingly reduced.)

Gay \& Lambert (2000) determined the Mg isotopic ratios in 20 stars 
in the metallicity range $-1.8 <$ [Fe/H] $< 0.0$ with the aim of testing
theoretical predictions (e.g.~Timmes et al.~1995). These data exhibit a large dispersion
in  the $^{25,26}$Mg/$^{24}$Mg isotope ratio.  Hence, 
the case for a low $^{25,26}$Mg/$^{24}$Mg ratio at low [Fe/H] was, perhaps,
 not yet unambiguously established, although their results indicated 
that $^{25,26}$Mg/$^{24}$Mg appears to decrease at low metallicity for normal stars.  
Nevertheless, based upon the large dispersion in these data,
one could neither make the case for a high nor low ratio of heavy Mg isotopes.   
However, indirect evidence for Mg/Al-cycle
 element production in low-metallicity stars
 is suggested  by the observations of  Denissenkov et al.~(1998), 
who found substantial Al enhancements in globular-cluster giants 
at the expense of Mg abundances.

Although many of the stars studied by Gay \& Lambert (2000)
 were found to have Mg isotopic  abundance ratios somewhat higher than 
predicted, even the ``peculiar" stars which show enrichments in 
$^{25,26}$Mg do not have abundance ratios substantially above Solar. 
Moreover, it should be noted that Timmes et al.~(1995) point out
that their model underestimates the $^{25,26}$Mg by around a factor of 2 at a metallicity
of [Fe/H] $= -1$, and that this discrepancy may indicate that
 another source of magnesium isotopes is in operation at low metallicity.

Based on the galactic chemical evolution models available at that time (e.g. Timmes et al.~1995),
the adoption of Solar isotopic Mg ratios by Murphy et al.~(2003a) 
in the MM method would appear to have been safe and conservative. 
Previous models seemed to indicate that the $^{24}$Mg:$^{25,26}$Mg ratio 
was higher than 79:21 in the past, and it was shown that a 
composition more rich in $^{24}$Mg only strengthens the case for the evolution of $\alpha$.  
However, as we discuss in more detail below, 
increasing the abundances of the heavier Mg isotopes would yield 
a larger multiplet splitting from isotope shift effects. This would
imply an apparently higher value  for  $\alpha$ (i.e.~closer to the present value).
Indeed, raising the heavy isotope concentration to $^{24}$Mg:$^{25,26}$Mg = 63:37 
could remove the significance of the signal for  a time-varying the fine-structure constant. 

In support of this possibility, a new study of Mg isotopic abundances in stars in the 
globular cluster NGC 6752 (Yong et al.~2003) affects the case for assuming a Solar
isotopic composition. 
This study looked at 20 bright red giants  with an assigned metallicity of
[Fe/H] = -1.62. 
Since globular clusters may be the remnants of early galactic chemical evolution,
their abundances might be representative of the abundances to be found in the
DLAs of interest in the present study.
 The observations  of Yong et al.~(2003) show
a considerable spread in the Mg isotopic ratios which range from
$^{24}$Mg:$^{25}$Mg:$^{26}$Mg = 84:8:8 (slightly poor in the heavies)
to 53:9:39 (greatly enriched in $^{26}$Mg). 
Of the 20 stars observed, 15 of them show $^{24}$Mg fractions of 78\% 
or less (i.e.~enriched in heavy isotopes relative to Solar), and 7 of them show fractions of 70\% 
or less with 4 of them in the range 53 - 67 \%. 
This latter range, if representative of the abundances in the DLA systems,
is sufficiently low to have substantially affected any determination 
of $\alpha$.  A previous study (Shetrone 1996) also found unusually high abundances of 
the heavy Mg isotopes in  giant stars in the globular-cluster M13.  
For this system, $^{24}$Mg:$^{25,26}$Mg  as low as 50:50 
and even 44:56 was found.  Such values exceed the necessary condition to account for 
the apparent variation in the fine structure constant 
in the low redshift sample of Murphy et al.~(2003). 

Other chemical evolution models of low-metallicity systems also suggest that 
there may be a missing contributor to the cosmic Mg abundance. 
Standard chemical evolution models were compared to observations of DLAs 
 by Fenner et al.~(2003; 2004).  These studies indicate  that 
supplemental sources of Mg beyond massive stars are needed.  
They concluded that, although Mg is dominated in the present-day ISM by SN ejecta,
SNe probably played a lesser role to that of         IM         stars in earlier epochs.  

It should be emphasized that the $\alpha$ derived from the MM method 
is sensitive to the isotopic ratio of Mg and not the total abundance.  
The missing Mg from chemical evolution models and abundance measurements
points to another source, 
one which played a dominant role in the earliest epochs of star formation. 
In our models, this source is derived from AGB stars.

\section{Sensitivity of the MM method to Isotopic Shifts of Mg}
Before we describe the specifics of the chemical evolution model, 
it is necessary to illustrate how the heavy isotopes affect 
the determination of $\alpha$ from the MM method. 
Normally, the MM method would compare the line shifts 
of the species particularly sensitive to any real change in $\alpha$ 
to one with a comparatively minor sensitivity (referred to as an anchor).  
For the low redshift sample, which consists of 74 out of 128 measured
 absorption systems [and the most significant evidence for variation
in the Murphy et al.~(2003a) data as well as 
all of the data in the recent analysis of Chand et al.~(2004)], Fe lines
 are compared to Mg lines (which serve as the anchor).
 For our purposes, the sensitivity of the MM method to the fine structure constant
 can be 
roughly approximated by subtracting the wavenumber shift of Fe from Mg
\begin{eqnarray}
\Delta\omega_{Mg} - \Delta\omega_{Fe} = q_{eff}X =  q_{eff}2{\delta \alpha \over\alpha} ~~,
\end{eqnarray}
 where $\Delta\omega_{i}$ is the difference between
the observed wavenumber of line $i$ and
the laboratory value. The quantity $X$ relates to the change in $\alpha$,
$X \equiv (\alpha_z/\alpha_0)^2 - 1\approx  2 {\delta\alpha\over\alpha}$,
where the approximation is adequate for the small variations
of interest here.  
While the true method simultaneously fits ratios of Fe lines to Mg lines, 
the effects of isotope shifts
can be estimated by taking the average trend as is described below.
The quantity $q_{eff}$ is the difference between the average value 
of q for Mg and that for Fe. From the q values given in Murphy et al.~(2003a), 
we obtain $q_{eff}\simeq 1280 \pm 150$ from an average of the 74 low-$z$
 absorption systems. 

The same wavenumber shifts can be accomplished 
from the isotopic shift (IS) of Mg alone (since the IS for Fe is small)
\begin{eqnarray}
\Delta\omega_{Mg} - \Delta\omega_{Fe} = \omega_{A'} - \omega_{A}~~.
\end{eqnarray}
The IS depends sensitively on the field shifts, 
specific mass shifts, and normal mass shifts that in turn 
depend upon the nuclear charge configuration.  Hence, 
it must be determined experimentally.  
Accordingly, Berengut, Dzuba \& Flambaum (2003) have experimentally 
deduced the coefficients of these shifts.
They also provided a relationship between isotopic abundances and frequency shifts,
\begin{eqnarray}
\omega_{A'} - \omega_{A}={1\over c}[(k_{nms}+k_{sms})({1\over {A'}}-{1\over A})+F\delta<r^2>^{A',A}]~~,
\end{eqnarray}  
where $A'$ is the mean atomic mass (enhanced with the heavier isotopes) and 
$A$ is the Solar atomic mass number  (for Mg, $A = 24.32$).  The quantity  $k_{sms}$ 
is the specific mass shift coefficient, while $k_{nms}$ is the normal mass shift.   
$F$ is the field shift, while  $\delta<r^2>^{A',A}$ is the difference in the mean square radius, 
and $c$ is the speed of light.  Incorporating the coefficients given in 
Berengut et al.~(2003), we can determine the corresponding 
effective $\alpha$-variation for the $A'$ corresponding to a specific isotopic composition
of Mg.
Recall that increasing the abundances of the heavier Mg isotopes yields a 
larger  value for the deduced $\alpha$,
 and a ratio of $^{24}$Mg:$^{25,26}$Mg = 63:37 
is sufficient to obviate the need for a time-varying fine structure constant.
Note  that the IS also depends on the relative ratio of $^{25}$Mg versus $^{26}$Mg.  
The heavy isotopic ratio of $(^{25}$Mg+$^{26}$Mg)/$^{24}$Mg $= 37/63 = 0.58$ quoted above  
still assumes a Solar ratio of $^{25}$Mg$/^{26}$Mg.

\begin{figure}[ht]
\begin{center}
\mbox{\epsfig{file=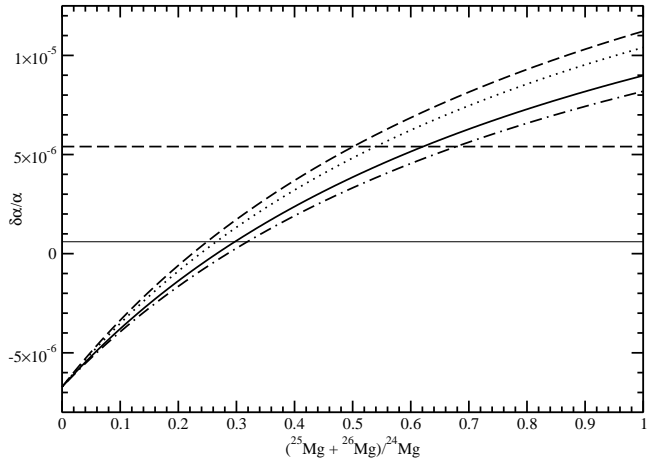,height=10cm,angle=270}}
\vskip .1 in
\caption{Effects of the Mg isotope shift.  Plotted is the change in 
deduced values of $\delta\alpha /\alpha$ 
relative to values obtained from an assumed Solar  Mg isotopic composition.
 The change in the deduced value is plotted  as a function of the
$(^{25}$Mg+$^{26}$Mg)/$^{24}$Mg 
isotopic abundance ratio.  The solid curve is the value deduced for a Solar
 ratio of $^{26}$Mg with respect to  $^{25}$Mg. 
The dotted (dashed) curve corresponds to $^{26}$Mg:$^{25}$Mg = 2:1 (3:1). 
The dot-dashed curve shows a sub-Solar ratio of
$^{26}$Mg:$^{25}$Mg = 8:10.  
The horizontal lines mark the value of $\delta\alpha /\alpha$
needed to compensate for the apparent shift in $\alpha$ determined 
by Murphy et al.~(2003a) (dashed) and Chand et al.~(2004) (thin solid).}   
\label{ratio}
\end{center}
\end{figure}

The change in the deduced values of  $ \delta \alpha / \alpha$ 
relative to those obtained when  assuming a Solar isotopic composition
(i.e.~the correction to
$\delta \alpha / \alpha$  due to non-Solar
values of the Mg isotopic abundances)
is shown in Fig. \ref{ratio}. 
Therefore, for Solar isotopic  composition [($^{25}$Mg+$^{26}$Mg)/$^{24}$Mg = 0.27],
 the value of $\alpha$ 
is the same as that obtained by Murphy et al.~(2003a) and Chand et al.~(2004).
  When only  $^{24}$Mg is assumed 
to be present, there is a shift if about $-0.67 \times 10^{-5}$, thereby
amplifying the signal they would have reported.  
Note that this shift is somewhat stronger than the 
shift quoted in Murphy et al.~(2003a) as we are using the more recent results 
of  Berengut, Dzuba \& Flambaum (2003) and Berengut et al.~(2003).

In contrast, when the ratio of heavies to 
$^{24}$Mg is larger, the deduced $\delta \alpha$ is smaller 
and can effectively cancel the effect obtained by Murphy et al.~(2003a). Using
our adopted  
mean value of $q_{eff}$, we find that the Mg isotopic ratio 
required to cancel the apparent result of Murphy et al.~(2003a) is 
($^{25}$Mg+$^{26}$Mg)/$^{24}$Mg = 0.62.  Figure \ref{ratio} also 
shows the dependence of $\delta \alpha$ 
on the ratio of $^{26}$Mg:$^{25}$Mg in addition 
to the total isotopic ratio.  If the ratio  $^{26}$Mg:$^{25}$Mg  were 2:1 or
3:1, the necessary total isotopic ratio, ($^{25}$Mg+$^{26}$Mg)/$^{24}$Mg, would be 
only 0.54 and 0.50, respectively. A sub-Solar ratio of $^{26}$Mg/$^{25}$Mg = 8:10
 would require the total isotopic ratio to be 0.68.  
Fig.~\ref{ratio} also illustrates that 
the ratio of $^{26}$Mg:$^{25}$Mg has a pronounced effect on the deduced value
of $\delta\alpha /\alpha$.  The deduced value of 
$\delta\alpha /\alpha$ increases by  $\approx 1.5  \times 10^{-6}$ even for
a Solar ratio of ($^{25}$Mg+$^{26}$Mg)/$^{24}$Mg = 0.27
just by varying the ratio of $^{26}$Mg:$^{25}$Mg  from 1:1 to 3:1.

\begin{figure}[t]
\begin{center}
\mbox{\epsfig{file=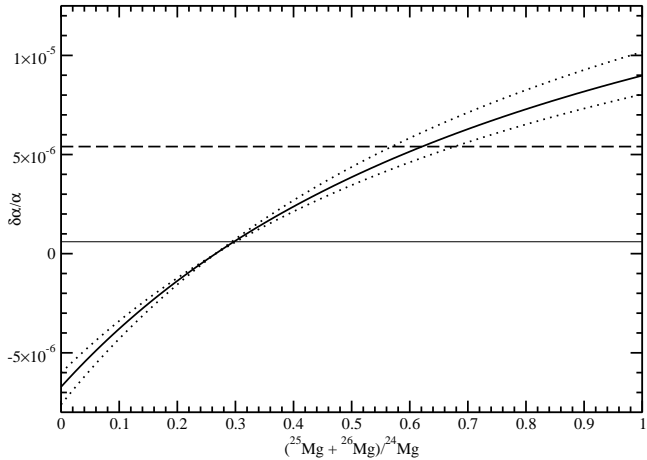,height=10cm,angle=270}}
\vskip .1 in
\caption{The shift in deduced values of $\delta\alpha /\alpha$ 
relative to values deduced from an assumed Solar  Mg isotopic composition
resulting from different values of $q_{eff}$.  The solid line is the shift
using the mean value of $q_{eff}=1280$.  
The dotted lines show the effect of the $1\sigma$ limits on $q_{eff}$, 
$\pm 150$, where the uncertainty is
dominated by the uncertainty of the $q_{Fe}$ given in Murphy et al.~(2003a).
The steeper (flatter) slope corresponds to  $q_{eff} = $ 1130 (1430).}
\label{ratio2}
\end{center}
\end{figure}

The  change in the deduced value of $\delta\alpha /\alpha$ 
is also sensitive to the value of $q_{eff}$.
While we have approximated the MM method with an effective average
from the low-z absorption systems, this value is somewhat uncertain. 
In  Fig.~\ref{ratio2}, we illustrate the sensitivity of the
deduced value for $\delta\alpha /\alpha$ to $q_{eff}$.
 Adopting a Solar ratio for $^{26}$Mg:$^{25}$Mg and  taking the 
$1\sigma$ range of $q_{eff}$, we obtain 
a mean value and uncertainty in the Mg isotopic ratio needed to compensate 
the apparent deviation in the fine structure constant reported in Murphy et al.~(2003a) 
of ($^{25}$Mg+$^{26}$Mg)/$^{24}$Mg = $0.62 \pm 0.05$.  Similarly, we infer
($^{25}$Mg+$^{26}$Mg)/$^{24}$Mg = $0.30 \pm 0.01$ for the Chand et al.~(2004) data. 
This uncertainty in $q_{eff}$ translates into 
an uncertainty in $\delta\alpha /\alpha$ of 
$\pm 0.14 \times 10^{-5}$ near a shift in
$\delta\alpha /\alpha$ near $5.4 \times 10^{-6}$.

While $\delta\alpha /\alpha$  is sensitive to the Mg isotopic composition
in the low redshift sample, the systems at $z \ge 1.8$  primarily depend upon Si. 
In Berengut et al.~(2003), it was determined that the Si$II$ lines are 
relatively insensitive to a heavy isotopic shift, 
although heavy  Si isotopes may be produced in an analogous
 fashion to the Mg isotopes  in AGB stars.  
Fully accounting for the $\delta\alpha /\alpha$ in the 
sample with $z > 1.8$ via isotopic abundances 
would require (Berengut et al.~2003)
 a conspiracy of several isotopic abundances.

\section{DLA chemical evolution with early AGB enrichment}

In order to explore the Mg isotopic evolution in DLAs, 
it is important to build a plausible chemical evolution model that 
is constrained by observations. Unfortunately, 
detailed abundances for most  of the systems observed in the Murphy et al.~(2003a) 
have not been reported.
Nevertheless, we can assume that their sample of DLA 
absorption systems have similar characteristics to those 
of measured DLA systems within the same redshift range.  
Systems  in the  redshift  range of interest here ($0.5 < z < 1.8$)
typically span a broad range of metallicities
with the mean value around [Fe/H] = -1.1 (Pettini 1999, Cen et al.~2003). 
Pettini (1999) also demonstrated that there
 is no discernible  evolution of metallicity 
in the redshift range of $z = 0.4 - 3.5$.  
This lack of metallicity evolution in DLAs is 
further supported by Prochaska \& Wolfe (2000, 2002). 
DLAs are distinguished by their high neutral hydrogen column densities; 
however, if a system were to produce large amounts of metals, 
it would also tend to ionize the neutral hydrogen gas.  
Alternatively, systems with high column densities 
in neutral hydrogen should also have reached the conditions for star formation.
 In fact, no observed DLA has a metallicity lower than 
[Fe/H]  $< -3$ (Cen et al.~2003). 
Hence, we confine our chemical evolution models to the broad 
range of metallicities observed in DLA systems.

In this section, we describe a chemical evolution model that can 
account for the observed variation in the fine structure constant.
We utilize recent Mg isotopic  yields from AGB stars. 
We then compare the model abundances to observed abundances in DLA systems
as a constraint on this interpretation of the MM results.  

For our purposes, a simple recalculation of the model of 
Timmes et al.~(1995) with and without the enhanced 
contribution from AGB stars is sufficient.  
This allows us to make a direct comparison with the conclusions 
of the previous authors.  We utilize the same metallicity-dependent yields of 
WW95 for Type II supernovae, 
using the same choices for
explosion energies as in Timmes et al. (1995), [i.e.~ model $A$ of
WW95  for $m = 11-25$ M$_\odot$,
and model $B$ for $m = 30,~35,$ and 40 M$_\odot$, which was taken as 
the upper limit of the
IMF.] We adopt the metallicity-dependent stellar lifetimes and the yields for
the stellar mass range from $ 6-7 M\odot$ given by Portinari, Chiosi, \& Bressan (1998).   
We supplement the IM yields of Marigo (2001) for stellar masses up to 
$5 M\odot$ (using the $\alpha=2$ mixing length parameter model) 
with the recent AGB yields of Karakas \& Lattanzio (2003) 
where the abundances of all three Mg isotopes 
are included.  

Although the yields for massive stars are available from zero to Solar metallicity,
we need to extrapolate the yields for IM stars below the lowest 
adopted metallicity model of $Z = 0.004$.
We keep the yields constant when below the lowest metallicity model.  
This approach is adequate for primary elements, but needs to be justified
for elements such as N and Mg, which are known to derive from secondary 
sources at high metallicity.  Although N is largely a secondary element 
at high metallicity, it is known (Pagel 1997) to be largely primary at low metallicity ($Z ^<_\sim 0.004$).  Hence, constant nitrogen
for $Z ^<_\sim 0.004$) seems justified.  Likewise, Mg yields can be
held constant, since it roughly scales with N. Obviously, these assumptions
leads to some uncertainty and needs to be checked with future calculations of IM stars at lower metallicity. 
   
In addition to extrapolationg to low metallicity, 
it was also necessary to interpolate in the mass
range of $ 7 - 12$  M$_\odot$ between the IM models and WW95 models.  
This leads to an uncertainty in the yields for that mass range.
A simple linear interpolation between 
the IM and WW95 models would 
be dominated by the 
larger yields of the Type II SNe. Also, stars in that mass range
at some point no longer experience the thermally pulsing  AGB phase.
Therefore, we adopt a more conservative approach whereby we 
interpolate the yields for both models to zero for 9 M$_\odot$ stars
in order to distribute the uncertainties in the yields between both
sets of yield models.  
We also adopt the Solar abundances of 
Grevesse \& Sauval (1998).   
 
The model of Timmes et al.~(1995) is based upon an exponential infall of primordial gas
 over a timescale of 4 Gyr, and a Schmidt law for the star formation rate. 
For our purposes, the efficiency of star formation can be modified to best account for 
the observed abundances in DLAs.  This is particularly true given that
they are most likely the progenitors of a wide range of galaxy morphologies. 

We assume instantaneous mixing with no outflow.  
Hence, the evolution of the mass fraction  $X_i$ of isotope $\it i$ can
be written,
\begin{eqnarray}
{d ({M_{g}(t) } X_i) \over dt} = 
  m_{CO} X_i^{Ia}R_{Ia}+\dot M_{g,i}
-B(t) {X_i(t)}  \nonumber \\
+ \int_{0.8}^{40} B(t - \tau(m)) \Psi(m) X_i^{S}(t-\tau(m)) dm \nonumber \\ 
+ \int_{2.5}^{7.0} B(t - \tau(m)) \Psi(m) X_i^{AGB}(t-\tau(m)) dm
\nonumber \\
\label{sigdot}
\end{eqnarray}
where, $B(t)$ (in units of M$_\odot$ Gyr$^{-1}$) 
is the stellar birthrate function at time $t$, 
$\Psi(m)$ is the initial mass function (IMF), 
$X_i^x$ is the mass fraction of isotope $i$ 
ejected by various sources specified by superscripts 
($x = S$ for normal evolution, 
$x = Ia$ for supernovae Type Ia ejecta, 
and $x = AGB$  for the supplement ejecta of Karakas \& Lattanzio(2003)). 
The lifetime of a progenitor  star of mass $m$
is denoted $\tau(m)$,  while $m_{CO}$ is the mass of the carbon-oxygen white-dwarf
progenitor of the Type Ia supernovae,  with $R_{Ia}$ as the rate of
Type Ia supernovae. The quantity $M_{gas}$ is the gas mass at time $t$. 
$\dot M_{g,i}$ represents the galactic infall rate of isotope $i$ 
(presumed to be primordial material). 
The third term  on the right hand side of Eq. \ref{sigdot} 
refers to the amount of isotope $i$ incorporated into new stars. 
The yields given by Karakas \& Lattanzio (2003) are for AGB stars up to 6.5 M$_\odot$.
Even at $m = 6.5$ M$_\odot$ the production of heavy Mg isotopes is still increasing
with mass.  Therefore, we extrapolate the Karakas \& Lattanzio (2003) yields
to 7 M$_\odot$.  Beyond that, the effective Mg yield is linearly interpolated 
to zero at 8 M$_\odot$ because such stars do not ignite He shell burning
under electron-degenerate conditions.  Neither do they
go through an AGB phase. Although our results are somewhat
sensitive to this extrapolation, a change in our result due to 
a different extrapolation could be compensated by a change in the 
AGB enhanced IMF.

 While the stellar evolution models of 
Karakas \& Lattanzio (2003) show robust AGB evolution beyond 5 M$\odot$, the
adopted CNO yields of Padova stellar models of Marigo (2001) and Portinari et al. 
(1998) do not exhibit AGB evolution beyond 5 M$\odot$.  Until self-consistent
yield models of AGB stars are developed that follow both Mg and CNO evolution,
we are limited to supplementing Mg evolution in the manner previously described.
Consequently, this approach may affect the direct coupling of the Mg to the
associated CNO abundances.     
  
In our model, the cosmic Type Ia supernova rate is given by 
the formulation of Kobayashi, Tsujimoto, 
\& Nomoto (2001). 
This rate incorporates a  minimum metallicity condition
before the operation of Type Ia supernovae 
 of  [Fe/H] $>-1.1$.
This is adopted as a necessary condition 
for the white dwarf progenitor to accrete effectively from the binary
companion.  
This condition was hypothesized by Nomoto et al.~(2003) 
based upon the fact that the
stellar wind velocity  from the binary companion 
is slower at low metallicity and also that the mass range
of white-dwarf progenitors is somewhat limited for the short timescales 
corresponding to low metallicity.

We propose a modest early enhancement of the IMF along the lines of 
Fields et al.~(1997; 2001). 
We present some theoretical and observational evidence for this enhancement below. 
Even though the early IMF is enhanced in the        IM         range, 
the usual normalization is still adopted
\begin{eqnarray}
\label{imfnorm}
\int \Psi(m)dm = 1
\end{eqnarray} 
We then write the star formation rate $B\Psi$ as
\begin{eqnarray}
\label{imf}
B(t)\Psi(m) & = &B_1(t) \Psi_1(m) + B_2(t) \Psi_2(m) \qquad     \\
= B_1 m^{-2.31}&+&(B_2/m) \exp{\biggl[-(\log{(m/m_c)}^2)/{(2 \sigma^2)}\biggr]}. \nonumber
\end{eqnarray}
The $\Psi_1(m)$ IMF accounts for a standard Salpeter distribution of stellar masses.
$\Psi_2(m)$ is an additional  log-normal component of 
stars peaked at $m_c$ M$_\odot$. 
The dimensionless width, $\sigma$, allows the mass distribution to extend 
across the entire    IM             range.  For the model of Ashenfelter,
Mathews, \& Olive (2004), $m_c = 5$ M$_\odot$ and $\sigma = 0.07$ was adopted.  
In the present work, a range
for these parameters is adopted consistent with the constraints deduced in
Adams \& Laughlin (1996).

For the normal stellar component, we parametrize the time dependence 
of the stellar birthrate function $B_1(t)$ as
\begin{equation}
B_1(t) = (1.0-e^{-t/\tau_1}) \epsilon_{SF}M_{tot}(t) \biggl[{M_{g}/{M_{tot}(t)}}\biggr]^2~~,
\label{b1}
\end{equation}
while for the        IM         component we similarly adopt 
\begin{equation}
B_2(t) = A_2 e^{-t/\tau_2}\epsilon_{SF} M_{tot}(t) [{M_{g} \over {M_{tot}(t)}}]^2~~.
\label{b2}
\end{equation}

This model thus contains an early burst of        IM         stars peaked at
$m_c$, with a spread of masses governed by the dimensionless width $\sigma$. 
The burst is exponentially suppressed on a time scale of $\tau_2$.  The $B_1$
component describes the standard quiescent star formation with a smooth
transition from the burst.  We note that the IM component obviously dominates
the mass recycle rate early on (depending on the values for the free
 parameters), but quickly gets diluted as the normal Salpeter component 
evolves. The coefficient $A_2$ in  $B_2(t)$ was 
adjusted to produce sufficient Mg isotope enhancement.  
Finally, $\epsilon_{SF}$ is the 
coefficient of star formation efficiency.  For reference, Timmes et al.~(1995) 
determined that the best fit for the Solar neighborhood was $\epsilon_{SF}$ = 2.8.
 
\begin{figure}[t]
\begin{center}
\mbox{\epsfig{file=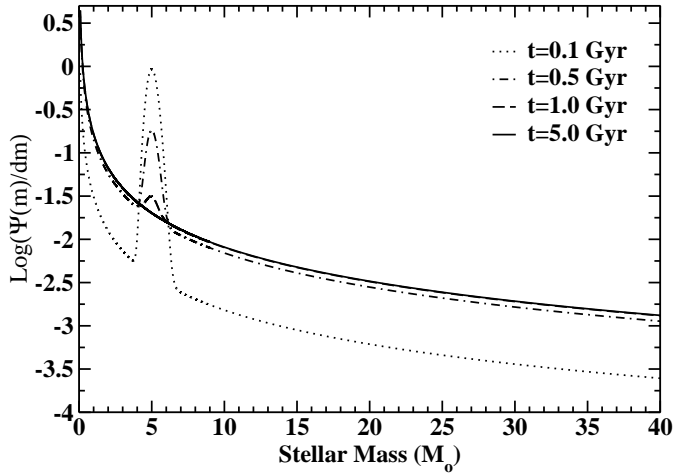,height=10cm,angle=270}}
\caption{Evolution of our adopted AGB-enhanced IMF.  
The log-normal component is centered at 5 $M_\odot$ with $\sigma = 0.07$
 as in our model $1$.
This enhancement, however, quickly evolves to a standard Salpeter IMF. 
The values of $\tau_1$ and $\tau_2$ are 0.5 and 0.2 Gyr, respectively. 
}   
\label{fig:imf}
\end{center}
\end{figure}

It should be noted that this coefficient gets normalized with the IMF, 
so that it only affects the abundance ratio when the 
log-normal enhancement is competing 
with the normal power law component of the IMF. The normalized IMF is shown at
four different times in Fig. \ref{fig:imf}.  
The sensitivity of our results to each of 
the parameter choices in Eqs. (\ref{imf}) - (\ref{b2}) 
will be discussed in section 6 below.

In addition to the overall star-formation efficiency, $\epsilon_{SF}$,
the model consists of two parameter categories: the parameters that
 determine the total amount of        IM         stars above the standard
 IMF ($A_2, \tau_1,\tau_2$) and those that determine the distribution within
 the enhancement ($m_c,\sigma$).  The exponential decay of the 
 enhancement of IM stars  ensures that  a smooth transition into the standard IMF occurs.  
The combination of $A_2$ and $\tau_2$ determines the ratio of the two IMFs
while they are in competition.

\subsection{Arguments for an AGB-enhanced IMF at low metallicity}

To some extent, our adopted increase in the IMF for IM stars may simply be 
thought of as a
compensation for the uncertainties in the nucleosynthesis yields of
AGB stars.  Indeed, a very small change in burning temperature
or a more efficient dredge-up could
easily accomplish the required enhanced $^{25,26}$Mg yields.
Nevertheless, there are also reasons to expect that the early IMF 
would be considerably different than the present day IMF.
Hence, we adopt an enhanced AGB-star IMF as a reasonable parameterization.

Arguments for an enhanced IMF for IM stars are as follows:
with fewer metals initially, cooling in the dense protostellar cloud
is predominantly done by atomic hydrogen, which is less efficient.
Therefore, a more massive cloud  is required to 
cool and gravitationally collapse into stars.
The lower limit of the IMF depends sensitively
on the length scale of density fluctuations and turbulent fragmentation. 
These density fluctuations in the present-day interstellar medium (ISM) 
are largely the result of previous star formation, which was obviously absent 
in the first star formation epoch. Moreover,
these primordial clouds are at higher temperature in the low-metallicity ISM.
Therefore, they further inhibit 
the smaller length scales of the density fluctuations.  
Uehara et al.~(1996) set the lower mass limit of the extremely 
metal poor stars at $ 1.4 M_\odot$, which consequently implies that 
all metal-free stars have since evolved into white dwarfs or other remnants.  
Although a 1.4 M$_\odot$ cut-off does not necessarily imply 
a higher relative abundance of IM with respect to massive stars,
arguments of a more steeply sloped early IMF have 
also been proposed which would suppress massive stars relative
to the IM stars of interest here. For example,  Padoan \& Nordlund (2002)
suggest a steeper IMF slope  as due to weaker magnetic fields in molecular coulds
at high redshift.

A further constraint is provided by the models of Yoshii \& Saio (1986), 
who concluded, based on an opacity-limited fragmentation model, 
that a metal-free IMF should be peaked at $ 4 - 10 M_\odot$.
It should then decline steeply on both sides of the peak. 
Furthermore, competing models of a metal-free IMF have 
the peak at much higher mass (Hernandez \& Ferrara 2001) and can even allow for 
very massive objects (VMOs) to form (Susa, Uehara, \& Nishi 1996).   

Further potential constraints on an enhanced IMF for IM stars
exist from observations of stars in the Galactic halo, although we emphasize that
details of the local Galactic environment do not necessarily
apply to the Lyman-$\alpha$ systems of interest to the present study.
Even so, Adams \& Laughlin (1996) and Fields et al.~(1997) 
(see also Ryu, Olive \& Silk 1990)
delved into the form of the IMF that could match the MACHO
 data of gravitational microlenses in the Galactic halo.  They independently 
determined that the early IMF is tightly constrained 
to be in the mass range of $ 1<M_\odot<8 $.  
These        IM         stars would evolve into white dwarfs and populate the halo 
if they had formed early enough to evolve and cool. 
Specifically, they determined that the maximum of the log-normal 
distribution is around $2-3 M_\odot$ and is sharply peaked.
  
Regarding those earlier models that produce
a very strong enhancement of IM stars, it has since been argued that 
those models overproduce current constraints on halo white dwarfs
(Gibson \& Mould 1997;
Flynn, Holopainen \& Holmberg 2003; Garcia-Berro et al. 2004, Lee et al. 2004).
We note, however, that these MACHO IMFs are more significantly enhanced with 
IM stars
than the models considered here. While their goal was
to generate a significant fraction of the Galactic Halo in white dwarfs,
our intention is to produce enough Mg isotopes from IM stars to imitate a
variation in $\alpha$. 
Since this model quickly vanishes at early times
due to the exponential decay factor in equation (\ref{b2}), it only can
partially populate the Galactic Halo.  
An important consideration since the earlier work on the MACHO IMF models 
is that the estimated age of the universe has been reduced so that the
white dwarfs would have had less time to cool. A smaller estimated
age of the universe could be offset by a mass distribution that
is peaked at a higher mass.     
  
Furthermore, MACHO IMF models are probably still difficult to reconcile with
the C, N, and O  constraint (Gibson \& Mould 1997) and new observations of 
old white cool halo white dwarfs Flynn, Holopainen \& Holmberg (2003). 
  In spite of these difficulties with the halo MACHO models, we emphasize
that the present models for Lyman-$\alpha$ systems 
are not affected by these constraints, both because the present models produce
fewer white dwarfs than the MACHO models, and because the evolution of the Lyman-$\alpha$ systems is not necessarily
the same as that of the Galactic halo.

\subsection{Observational Evidence of AGB-heavy IMF at low metallicity}

There is a large body of evidence suggesting a strong 
early contribution from          IM         stars.  Indeed, 
ejecta from such AGB stars 
is needed to account for the observed abundances in both the
low-metallicity stars in or near the Milky Way and those deduced in DLAs.  
Abia et al.~(2001) give an overview of the abundances of 
the most extremely metal-poor stars in our galaxy and argue that an IMF 
peaked in the    IM             range of  $ 3 - 8 M_\odot $ is favored over an IMF 
populated by VMOs.  This argument is
  based on the observed  large number of low-metallicity halo stars with 
greatly enhanced C and N abundances. 
Intermediate-mass stars are particularly efficient at producing
very large [C,N/Fe] ratios found in $\sim$ 1/4 of the iron deficient stars 
(Norris, Beers, \& Ryan 2000; Rossi, Beers, \& Sneden 1999). 
The most likely source of these abundances is accretion from a binary AGB companion. 
Reinforcing this conclusion are the observations that s-process elements, especially Pb, 
appear even at the lowest of metallicities (Aoki et al. 2001; Sivarani, Bonifacio, \& 
Molaro 2003, Sivarani et al.~2004). Only in some cases  can 
the s-process enrichment be attributed
to Roche-lobe overflow from a companion.  In others, an alternative explanation may be necessary.    
Because massive stars pollute the ISM before    IM             stars of the same age, 
the s-process cannot ``outpace" the r-process except in  several special circumstances:  
The r-process elements may not have mixed efficiently into the ISM and cluster; 
or the local first epoch of star formation may have been dominated by        IM         stars
as described herein. 

It has also been argued that in order to account for the depletion of Deuterium, 
but still produce the 
observed white-dwarf population found the Galactic halo, a non-standard IMF 
is required (Fields et al.~1997; 2001). In particular, an early IMF dominated by
       IM         stars has the advantage of matching these characteristics
of galactic protodisks without the problem of metal over production from massive
stars. In fact, our star formation rate, 
Eqs. (\ref{imf}) - (\ref{b2}), is based on the IMF proposed in Fields et al.~(2001).

There are two obvious constraints on the magnitude of the 
enhancement in the AGB IMF at low metallicity.  These are 
carbon and nitrogen  production, and the implied Type Ia supernova rate.  
An early IMF enhancement has an insignificant 
effect on the Type Ia rate because of the adopted minimum metallicity requirement of 
Nomoto et al.~(2003). By the time the ISM reaches the minimum metallicity, 
the IMF is dominated by the Salpeter IMF. 
An early IMF enhancement would, however, inevitably produce more carbon and nitrogen.  
Indeed, recent abundance determinations in low-metallicity systems have 
already shown a significant enhancement in carbon and nitrogen. 
 
Two such systems are the low-metallicity ($Z \sim 0.004 $) globular clusters 47 Tucanae and M71,
which show a strong pollution of AGB ejecta early in their evolution, 
yet maintain a constant iron abundance (Briley et al.~2003; Harbeck, Smith, 
\& Grebel 2002). 
Many of the observed stellar abundances show C depletion correlated
with strong N enhancements.  
This trend is observed in both red giants and main-sequence turnoff stars, 
thereby making consistent internal contamination unlikely.   
Because globular clusters are sensitive to feedback from supernovae, 
we argue that the earliest star formation had to be comprised of mostly 
       IM         stars and not massive stars.  
Otherwise, the massive stars would contribute to the ISM first, and the C,N 
enhancements with respect to iron group would be diluted. 
Furthermore, SNe and the UV contamination by massive stars would 
suppress the star formation rate through feedback processes.
We note that this model does not contradict the possibility of a very 
early population of very massive stars which provide an effective
prompt initial enrichment (Z $\sim 10^{-3}$) and possibly reionize the Universe
at redshifts $z > 6$.

The Type Ia supernova rate does not depend upon our AGB
enhanced IMF because of the adopted minimal metallicity constraint on the efficiency of binary accretion.  Roughly 2/3 of the iron produced comes from Type II and the rest is from Type Ia, as is consistently implemented in other chemical evolution models.  
 Matteucci \& Recchi (2001) determined that
a metallicity condition on Type Ia supernovae varies from the observations
of the abundances in the Solar neighborhood.  Allowing for Type Ia supernovae  in our
model at the earliest metallicities would shift the abundance evolution
towards higher [Fe/H].  
While the total abundance of Mg does not impact the QSO absorption spectra, 
a comparison of the degree to which an AGB enhanced IMF is 
distinguished from the constant IMF is detailed below. 
Both theoretical and observational studies of low-metallicity systems seem to point 
to an IMF dominated by AGB stars and their associated ejecta, at least in some systems. 
Competing hypotheses for the early IMF such as a constant IMF or 
VMOs have advantages in certain situations but often must also be fine-tuned 
to account for the full gamut of observations.

\section{Results}
We next probe the consequences of our adopted  chemical evolution model on 
the production of the Mg isotopes and hence on the deduced  shift 
$\delta \alpha$ of the fine-structure constant.  
We compare these results to a ``standard'' 
model of chemical evolution (Timmes et al.~1995), which does not
include the yields of AGB stars.   This is listed as model 0 in Table 1.
We also compute the abundances of other elements and
compare them to observations of DLA systems.  It is particularly important
to consider nitrogen and carbon abundances because these are also ejected from the
       IM         stars responsible for Mg production.

We will begin with the model used in Ashenfelter, Mathews \& Olive (2004) (all models are summarized in Table 1).  
For this model (Model 1 in Table 1), we compute the abundances of N, N/Si, and C/O as a function 
of [Fe/H], [Si/H], and [O/H] respectively. We then explore the sensitivity of our results to
our assumed star formation rate. From this study, we obtain a set of
best-concordance parameters which reproduce the observed DLA abundances while still accounting for 
the data of Murphy et al.~ (2003a) (Model 2) or Chand  et al.~(2004) (Model 3) without 
varying the  fine-structure constant.

\begin{table}[htb]
\begin{center}
\begin{tabular}{|c|c|c|c|c|c|c|}\hline\hline
Model  & $m_c$    &     $\sigma$ & $\epsilon_{SF}$ & $A_2$ & $\tau_1$ & $\tau_2$ \\ \hline
0 &   -      &     -  & 2.8 &  -  &  -  &  -       \\ 
1 &  5.0     &    0.07& 1.9 & 8.9 & 0.5 & 0.2      \\ 
2 &   6.0      &  0.1 & 5.0 & 11.0 & 0 & 0.4         \\
3 &  6.0  &  0.1 & 5.0 & 3.6 & 0.1 & 0.3  \\
\hline\hline
\end{tabular}
\caption{Model Parameters
\label{tab:bbn}}
\end{center}
\end{table}

\begin{figure}[ht]
\begin{center}
\mbox{\epsfig{file=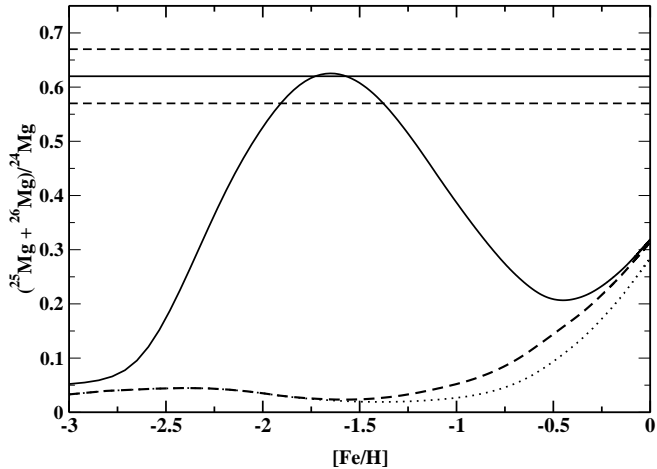,height=10cm,angle=270}}
\caption{The chemical evolution of the  $^{25,26}$Mg
isotopes relative to  $^{24}$Mg. The solid curve is our result based on 
Eq. (\protect\ref{imf}) using the AGB Mg yields of Karakas \&  Lattanzio (2003) and
the parameters of Model 1.  The dashed curve
is the result obtained when the burst of        IM         stars is excluded while still including 
the AGB Mg yields.  The dotted curve (Model 0) excludes both the AGB Mg yields and the
enhanced IMF.  The horizontal lines indicate the range for the ratio of $^{25,26}$Mg/$^{24}$Mg necessary to explain the shifts seen in the
 many-multiplet analysis as described in Section 4. }   
\label{old}
\end{center}
\end{figure}

As shown in Ashenfelter, Mathews \& Olive (2004), the birthrate function, Eqs. (\ref{imf}) - (\ref{b2}),
provides for a large enhancement in the production of $^{25}$Mg and $^{26}$Mg due to the
copious yields for these isotopes in IM stars at low metallicity.  
This result is shown in Fig. \ref{old}.   The 
evolution of the heavy Mg
 isotopes can be explained as follows:
Initially, the production of  $^{25,26}$Mg in the ejecta of        IM        
stars is delayed by their relatively long lifetime (compared to massive stars).
Initial contributions to the chemical enrichment of the DLA interstellar medium 
comes from the most massive and shortest lived stars.  In this model, the burst of 
     IM           stars begins to produce  $^{25,26}$Mg at [Fe/H] $\ga -2.5$.
At this stage, during the        IM         star-formation burst, $^{25}$Mg and $^{26}$Mg are
efficiently produced relative to $^{24}$Mg as per the yields of \ Karakas \&  Lattanzio (2003). 

At higher metallicity, the ejecta from the standard population of (massive) stars
(which is depleted in $^{25,26}$Mg) begins to dilute the ratio relative to $^{24}$Mg. 
This accounts for the decline from local maximum in $^{25,26}$Mg/$^{24}$Mg around
[Fe/H] $^>_\sim -1.5$. At late times, the impact of the early generation of      IM          
stars is diluted by the ejecta from subsequent generations of stars.
The dashed curve excludes the burst of IM stars, while the dotted curve  excludes
the AGB yields as well as the        IM         component. 
This latter curve essentially reproduces the result of Timmes et al.~(1995), which we refer to 
as Model 0.
We note that the new AGB yields were also included in the chemical evolution model of
Fenner et al.~(2003) who utilized a normal stellar IMF.
Their results are  similar to that shown by the dashed curve in Fig. \ref{old}.  
While these results show higher
abundances of $^{25,26}$Mg relative to $^{24}$Mg than that given by the dotted curve, 
they are not high enough to account for the apparent variability in $\alpha$.

\begin{figure}[ht]
\begin{center}
\mbox{\epsfig{file=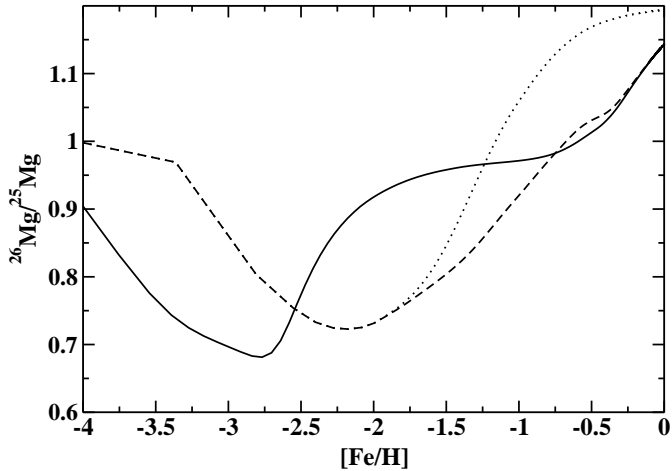,height=10cm,angle=270}}
\caption{The chemical evolution of the ratio of $^{26}$Mg/$^{25}$Mg. 
The curves correspond to the chemical evolution models as described in 
Fig. \protect{\ref{old}}.  The Solar ratio is 1.1. 
}   
\label{mg26vs25}
\end{center}
\end{figure}

As discussed in section 4, the degree to which the heavy isotopes of Mg can effectively shift
the value of the deduced fine-structure constant, depends sensitively on the  $^{26}$Mg/$^{25}$Mg
ratio.  This evolution of this ratio for Model 1 is shown in Fig. \ref{mg26vs25}. As one can see,
the $^{26}$Mg/$^{25}$Mg ratio remains slightly less than 1 for most of the metallicity range.  
This is due to the large enhancement 
of $^{25}$Mg over $^{26}$Mg in $\sim 6 M_\odot$ stars. The fact that this ratio is 
slightly less than Solar, requires a slightly larger excess of $^{25,26}$Mg relative to $^{24}$Mg
to explain the shift in $\alpha$.

\begin{figure}[t]
\begin{center}
\mbox{\epsfig{file=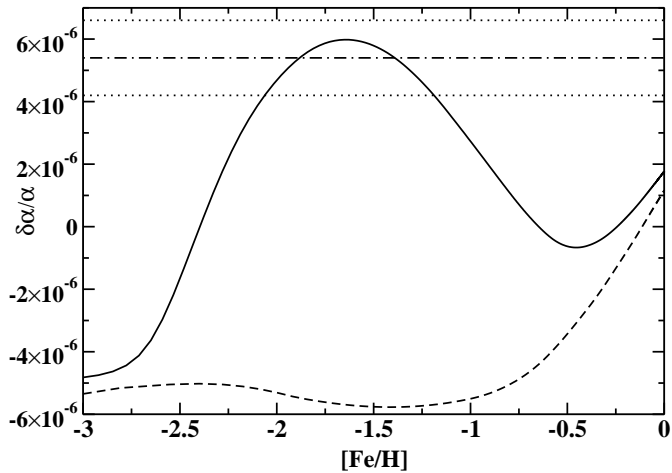,height=10cm,angle=270}}
\caption{Shift in the fine-structure constant due to isotopic abundance variations in Mg.  
The shift in $\delta\alpha /     \alpha$ relative to an assumed Solar Mg composition
 is shown for both models with (solid) and without (dashed) an AGB-enhanced IMF. 
The shift needed to compensate the value ascertained by Murphy et al.~(2003) is given by
the horizontal dot-dashed line with the $\pm 1\sigma$ errors given by the horizontal dotted lines.  
}   
\label{alfe}
\end{center}
\end{figure}

The most vital comparison we must make involves the determination of the effective $\alpha$ 
variation in our model as a function of metallicity.
Using the results of section 4, shown in Fig. \ref{ratio}, we
can determine the shift in $\delta \alpha$ based on the calculated Mg isotopic
abundances.  This result (for Model 1) is shown in Fig. \ref{alfe}.  
As one can see, the existence of  the enhanced isotopic abundances of $^{25,26}$Mg at a metallicity of 
[Fe/H] $\sim -1.5$ implies no true variation in the fine-structure constant.
 DLA  systems like the ones
 used to measure $\delta\alpha /     \alpha$ span a wide range of metallicities,
$-1.75 \le$ [Fe/H] $\le -0.75$ with an average value of [Fe/H] $\sim -1.1$ (Pettini 1999).
By construction, this coincides with area of maximum effect from isotopic 
variations of Mg in our model. The horizontal lines show the shift in $\alpha$ needed
to compensate for the variation deduced by Murphy et al.~(2003a).

\begin{figure}[ht]
\begin{center}
\mbox{\epsfig{file=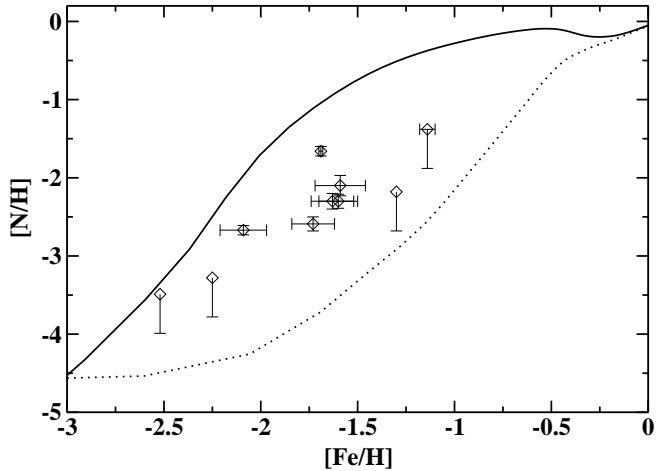,height=10cm,angle=270}}
\caption{Comparison of nitrogen evolution to observations in DLA systems.  
With an AGB enhancement (solid line), an excess of [N/H] is produced, 
while  a model without an AGB enhancement (dotted line) underproduces nitrogen.  
These data were taken from various abundance measurement of individual  DLA systems (Centurion et al 2003; Pettini et al 2002; D'Odorico \& Molaro 2004; Molaro et al.~2001; and Levshakov et al.~2002). 
}   
\label{fig:n}
\end{center}
\end{figure}

\begin{figure}[ht]
\begin{center}
\mbox{\epsfig{file=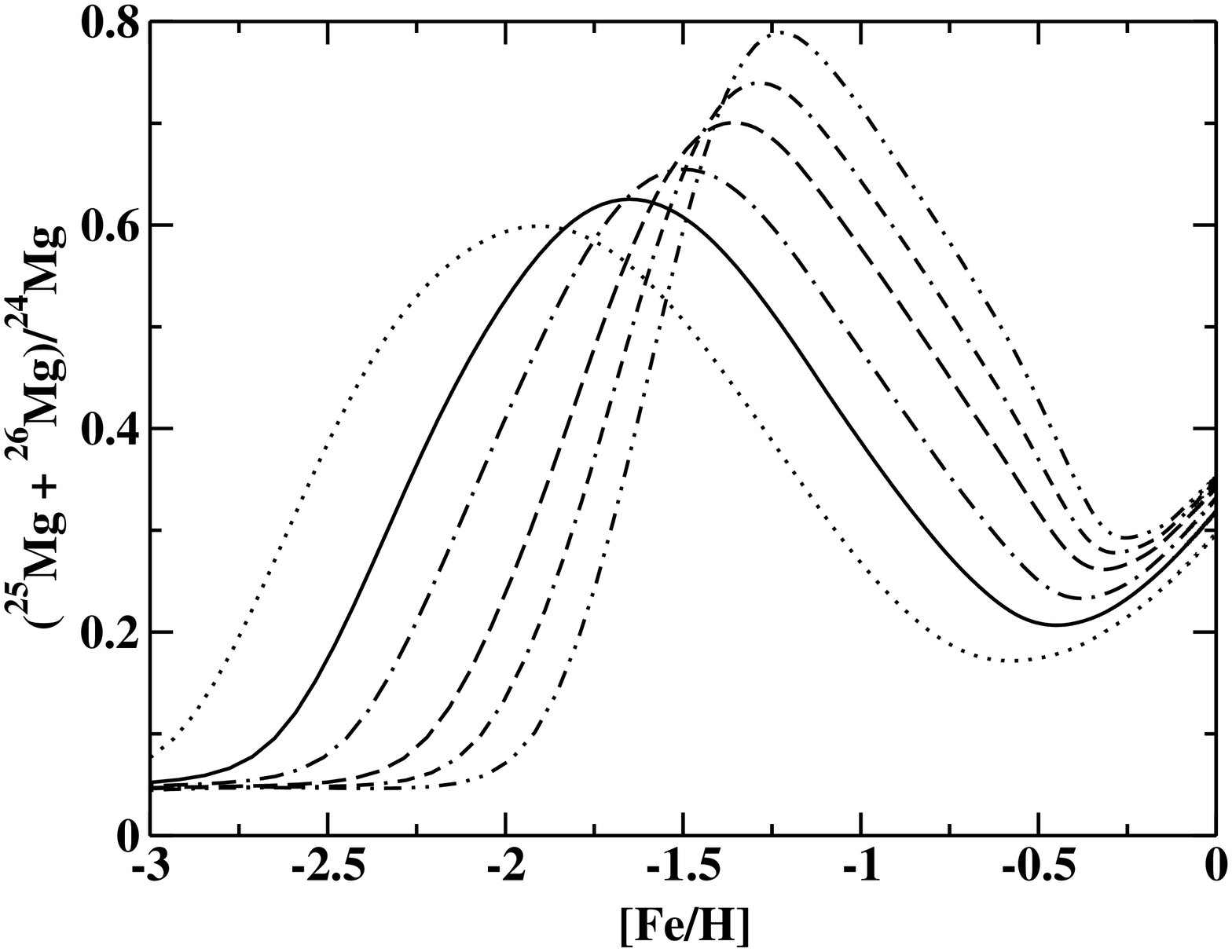,height=9cm,angle=270}
\epsfig{file=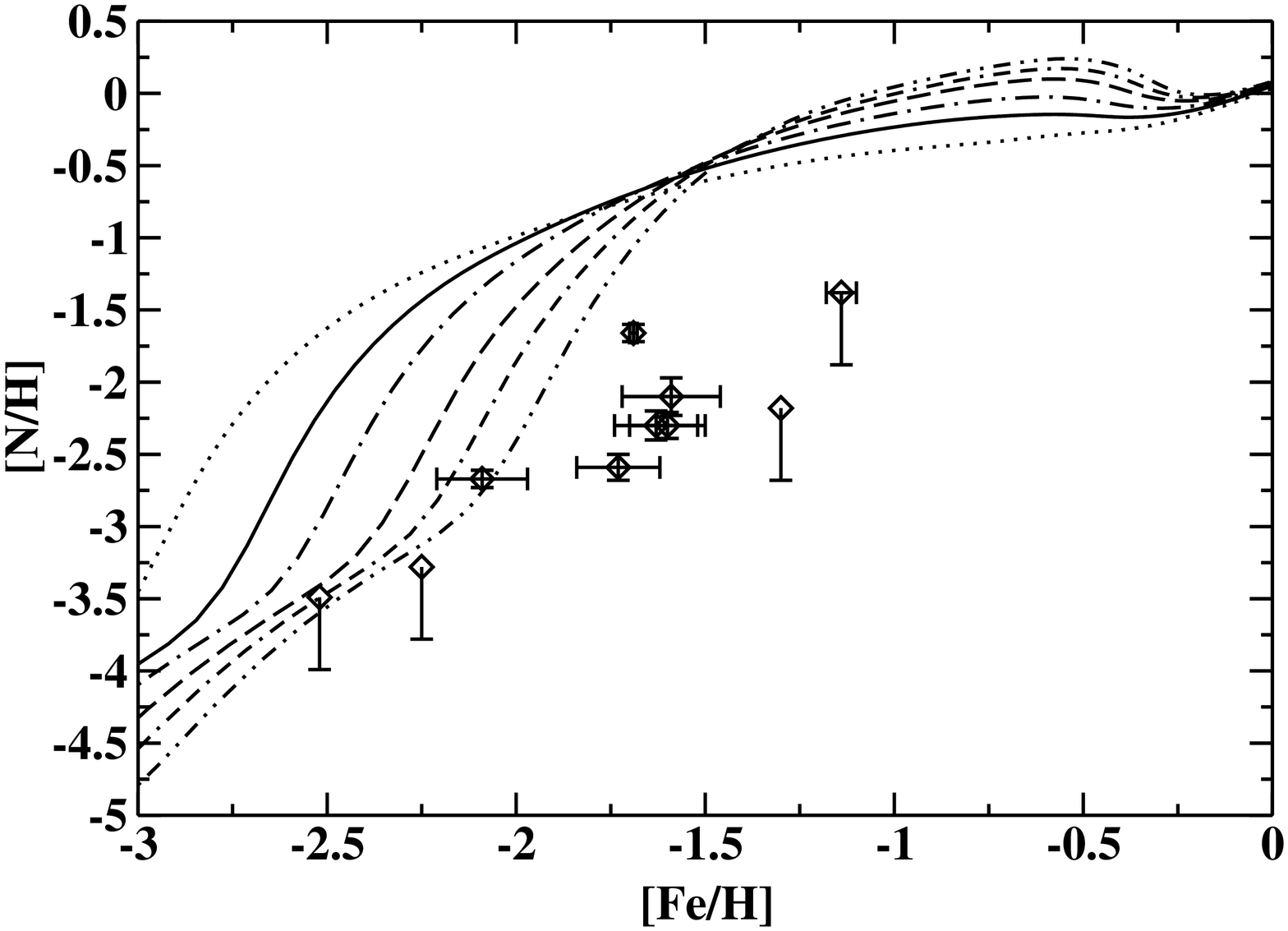,height=9cm,angle=270}}
\caption{Dependence of the magnesium isotopic ratio (left)  and the nitrogen abundances (right) on the 
coefficient of star formation efficiency, $\epsilon_{SF}$.  All 
parameters were kept constant except  $\epsilon_{SF}$, which had the values
of 1 (dotted), 1.9 (solid), 3 (dot-dashed), 5 (dashed), 7 (dot-dash-dashed), and 10
(dot-dot-dashed).  
}   
\label{stareff}
\end{center}
\end{figure}

\begin{figure}[ht]
\begin{center}
\mbox{\epsfig{file=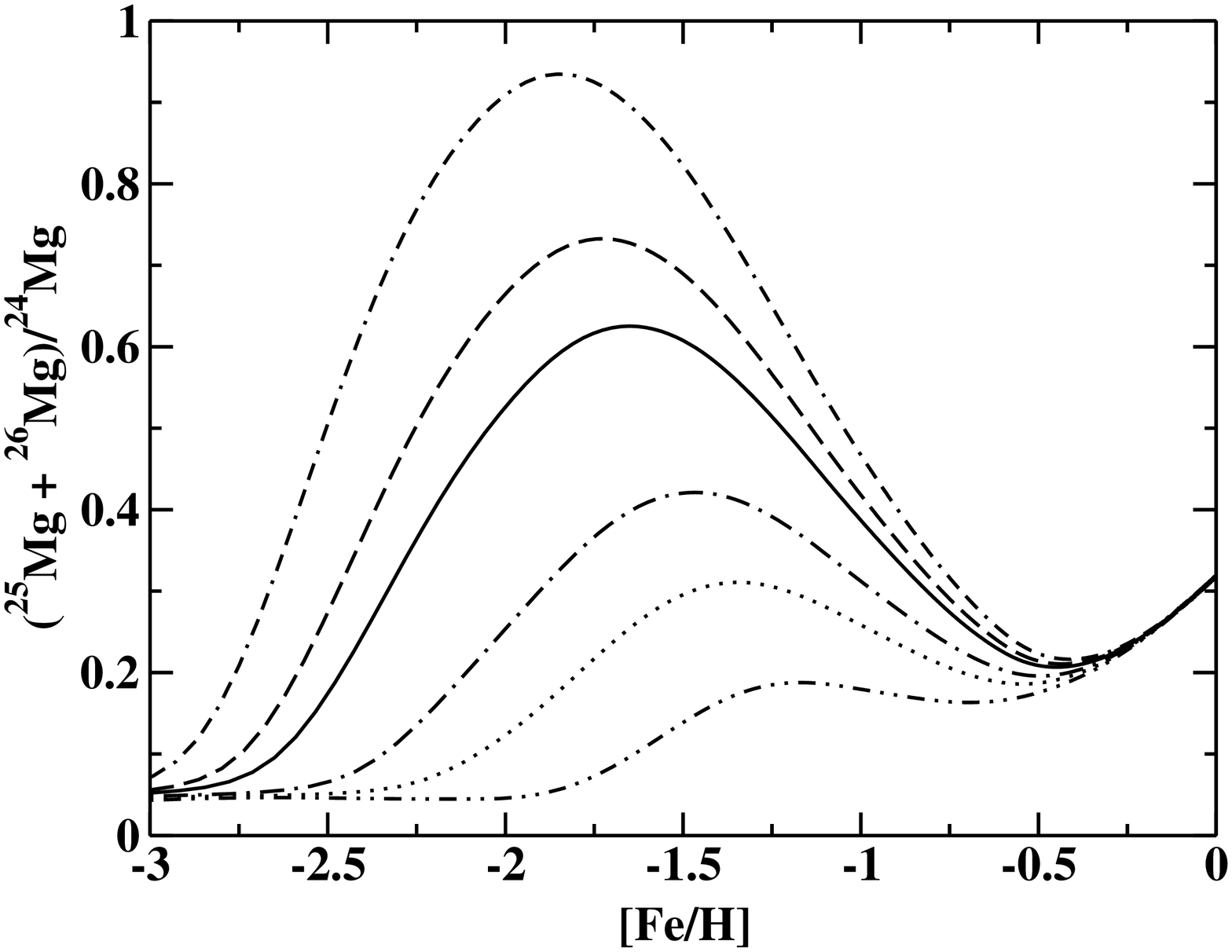,height=9cm,angle=270}
\epsfig{file=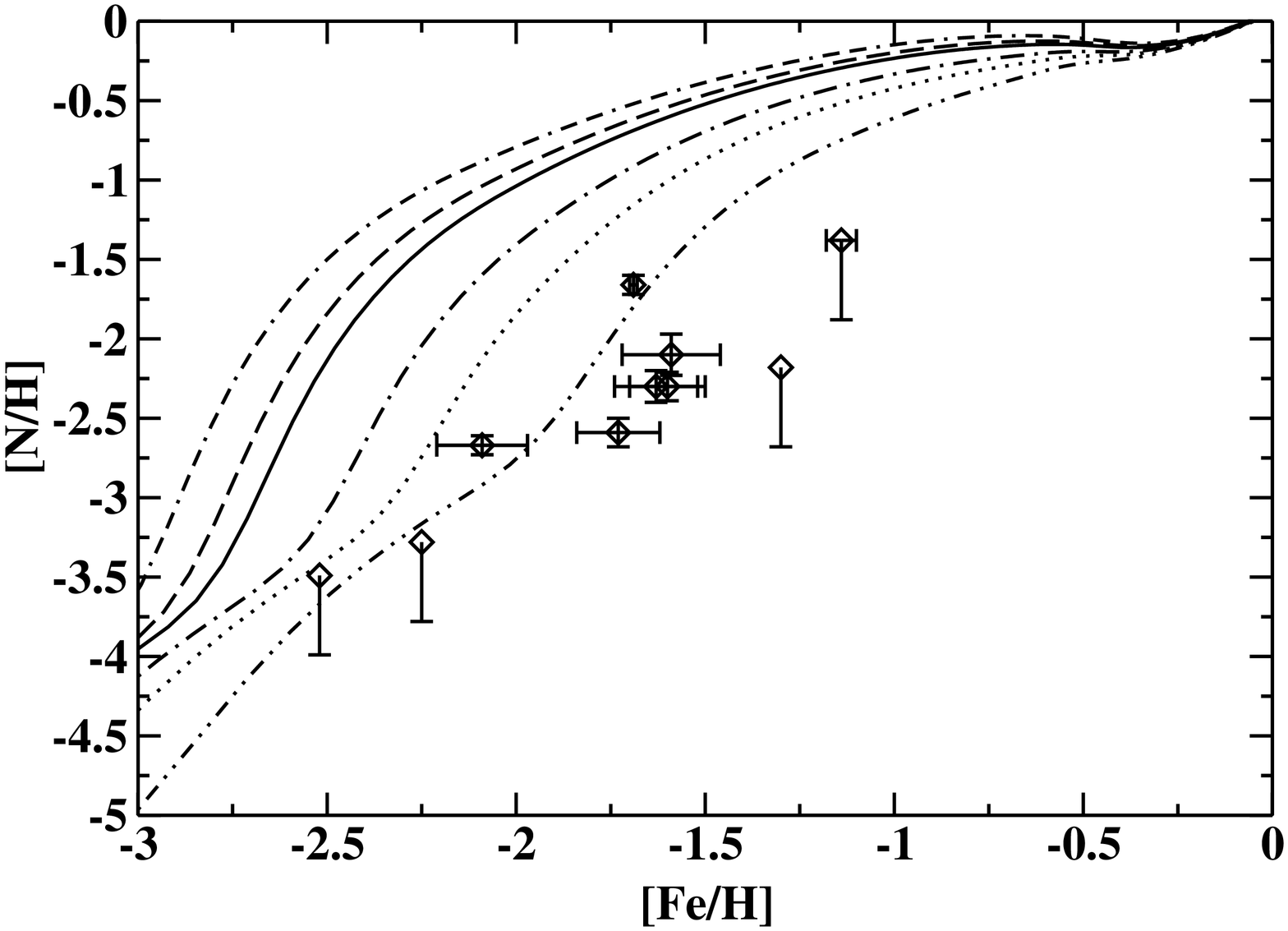,height=9cm,angle=270}}
\caption{Dependence of the magnesium isotopic ratio  and nitrogen abundance on the 
coefficient of the IM component of the IMF, $A_2$.  All 
parameters were kept constant except  $A_2$, which had the values
of 1 (dot-dot-dashed),  3 (dotted), 5 (dot-dashed), 8.9 (solid),  11
(dashed), and 15 (dot-dash-dashed).  
}   
\label{a2}
\end{center}
\end{figure}

\begin{figure}[ht]
\begin{center}
\mbox{\epsfig{file=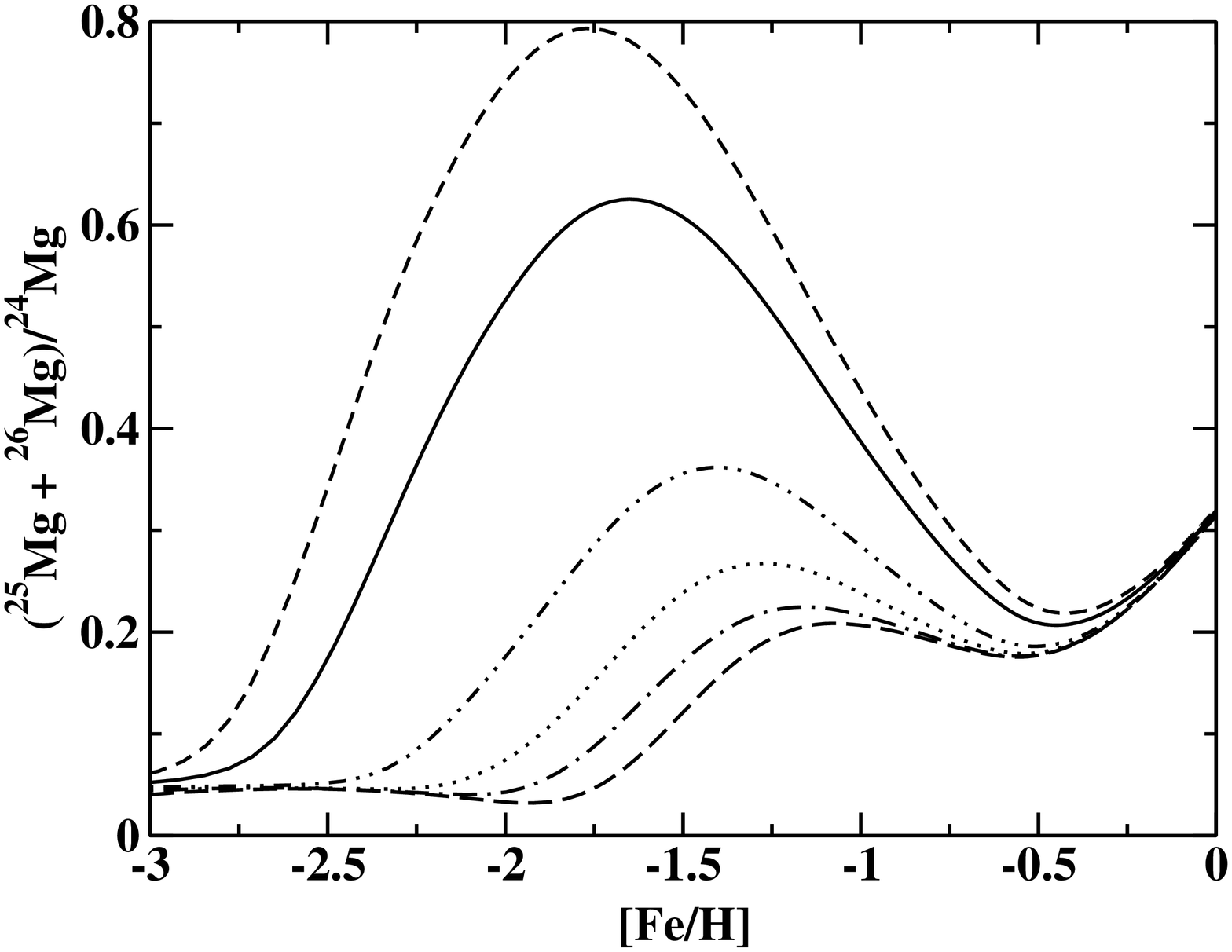,height=9cm,angle=270}
\epsfig{file=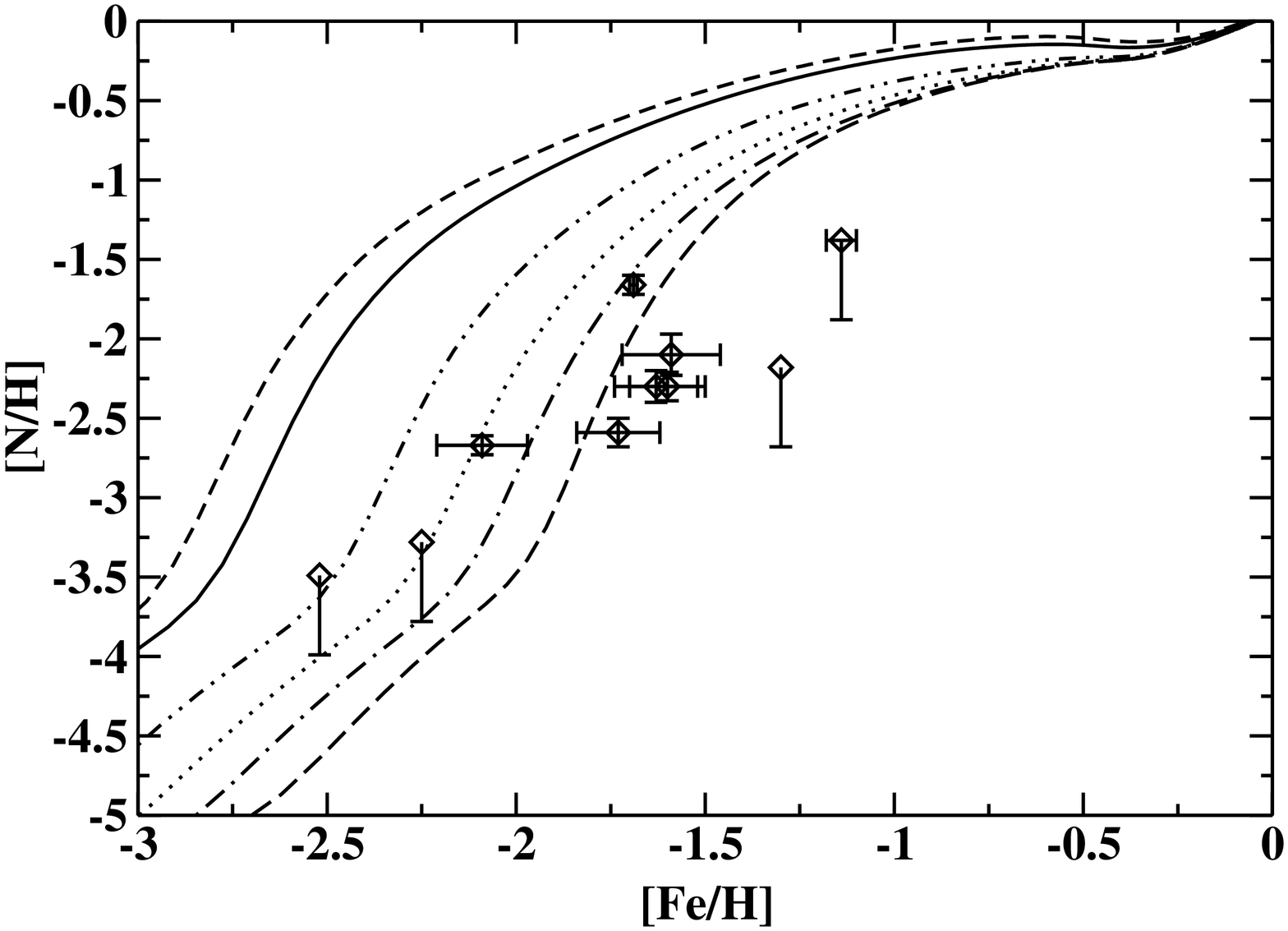,height=9cm,angle=270}}
\caption{Dependence of the magnesium isotopic ratio (left) and nitrogen abundance (right) on the 
turn-on time-scale of the standard component of the IMF, $\tau_1$.  All 
parameters were kept constant except  $\tau_1$, which had the values
of 0 (long-dashed),  0.05 (dot-dashed), 0.1 (dotted), 0.2 (dot-dot-dashed),  0.5
(solid), and 0.7 (short-dashed).  
}   
\label{t1}
\end{center}
\end{figure}

\begin{figure}[ht]
\begin{center}
\mbox{\epsfig{file=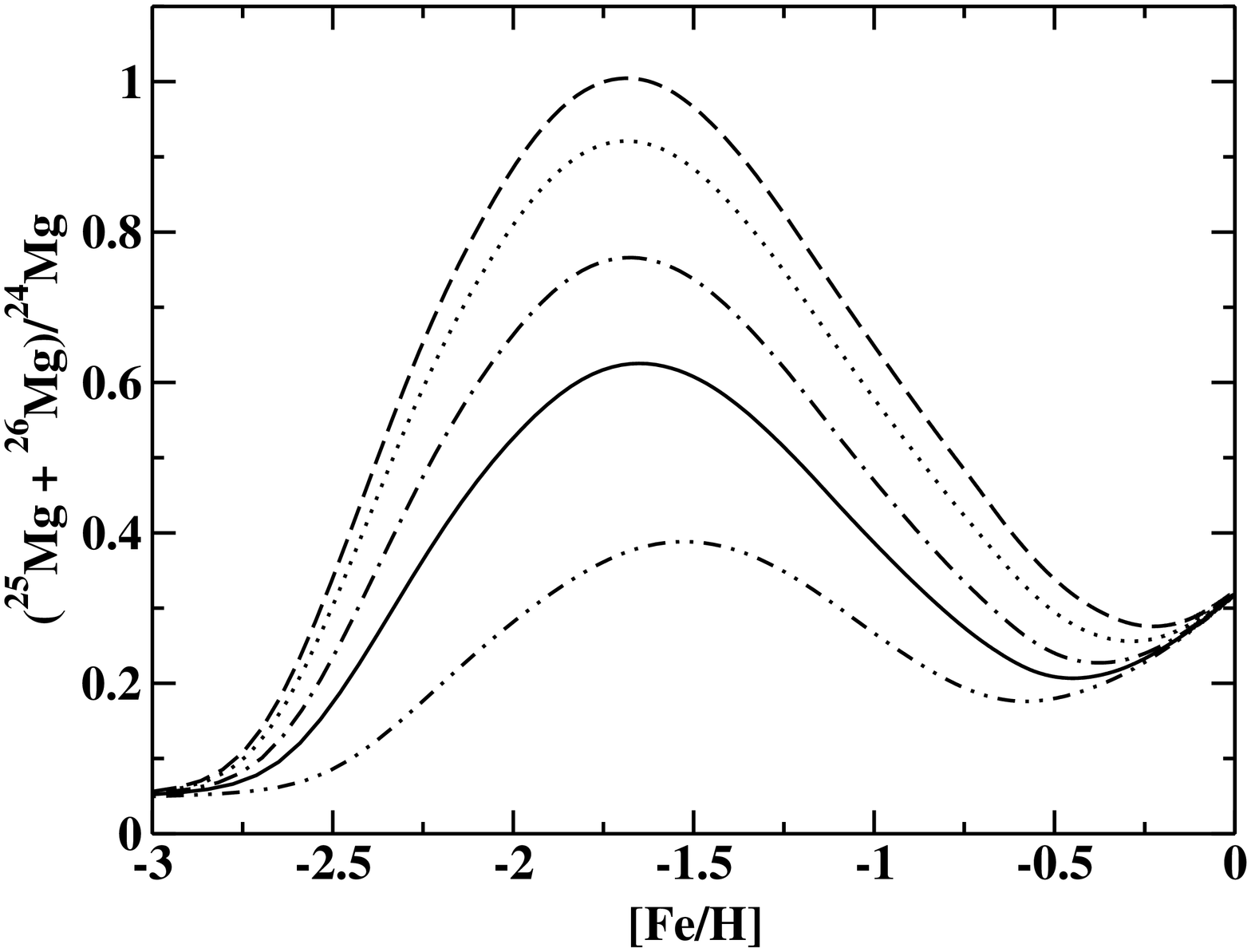,height=9cm,angle=270}
\epsfig{file=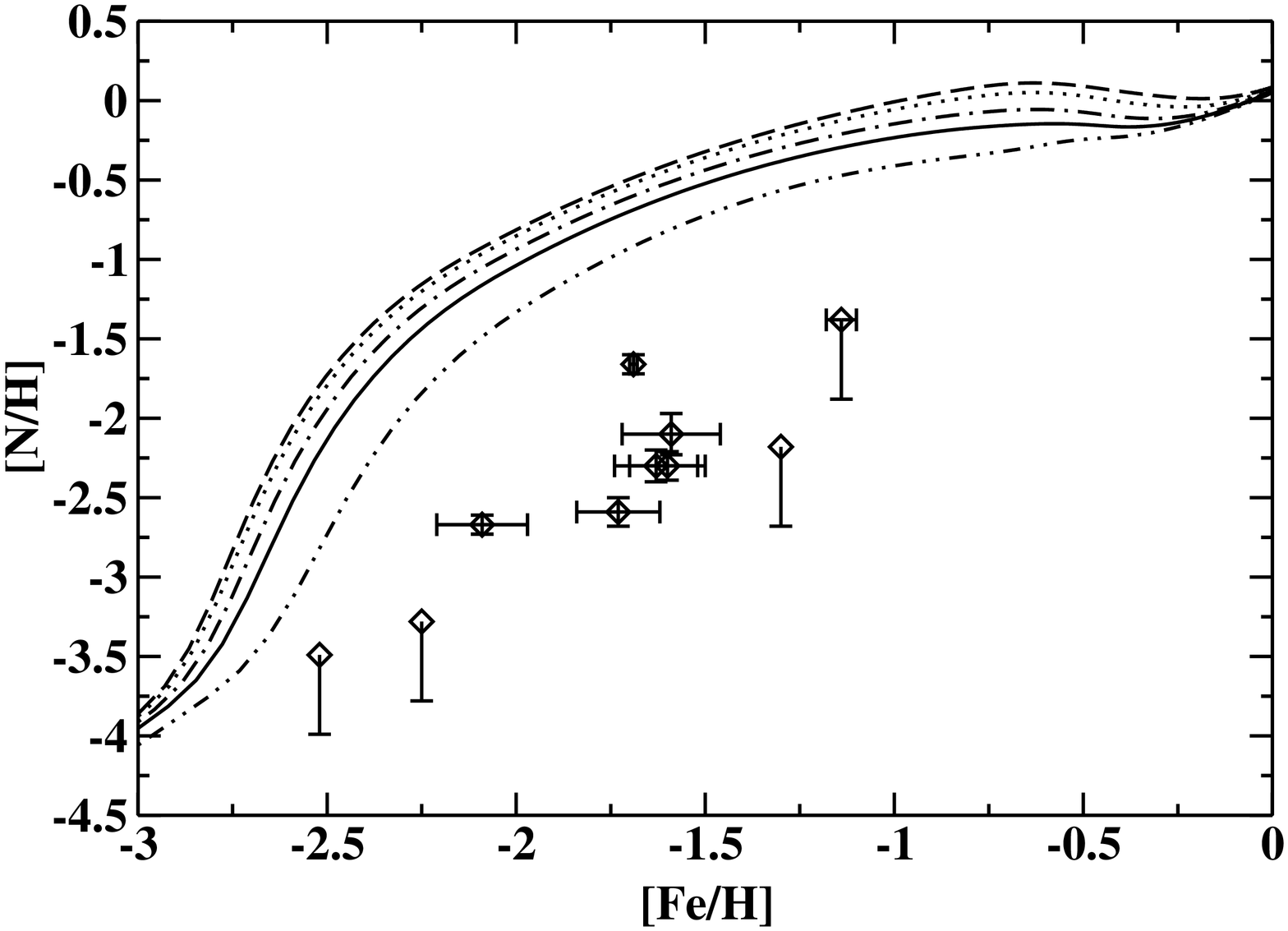,height=9cm,angle=270}}
\caption{Dependence of the magnesium isotopic ratio  and nitrogen abundance on the 
decay time-scale of the IM component of the IMF, $\tau_2$.  All 
parameters were kept constant except  $\tau_2$, which had the values
of 0.1 (dot-dot-dashed),  0.2 (solid), 0.3 (dot-dashed), 0.5 (dotted),  and 0.7 (dashed).  
}   
\label{t2}
\end{center}
\end{figure}

\begin{figure}[ht]
\begin{center}
\mbox{\epsfig{file=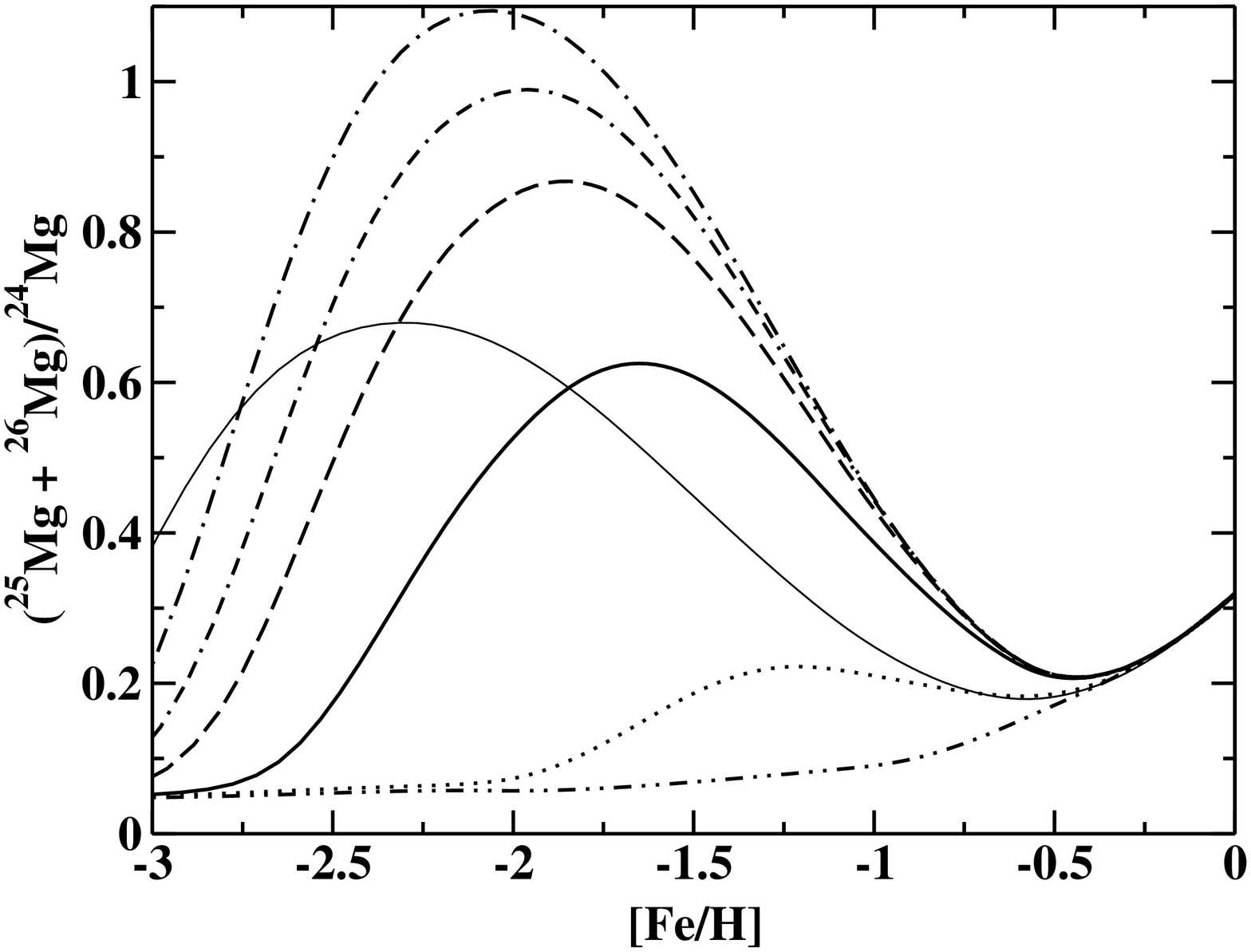,height=9cm,angle=270}
\epsfig{file=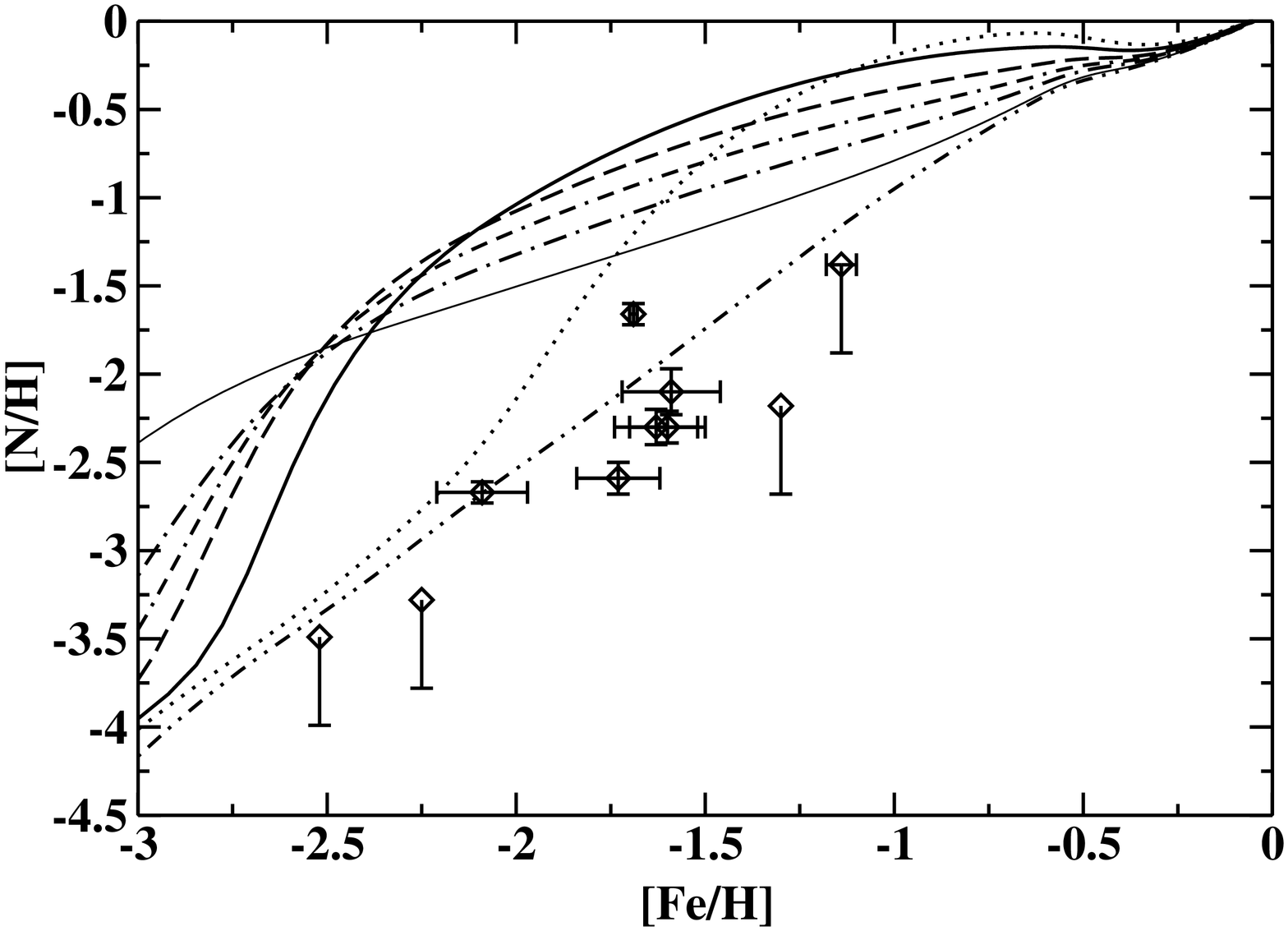,height=9cm,angle=270}}
\caption{Dependence of the Mg isotopic ratio (left)  and nitrogen abundance  (right) on the value of $m_c$.
 All parameters were kept constant except  $m_c$, which had the values
of 2.0 (dot-dot-dashed),  4.0 (dotted), 5.0 (thick solid), 5.5 (dashed),  5.75 (dot-dash-dashed),
6.0 (dot-dashed), and 7.0 (thin solid).    
}   
\label{mcmodel1}
\end{center}
\end{figure}

\begin{figure}[ht]
\begin{center}
\mbox{\epsfig{file=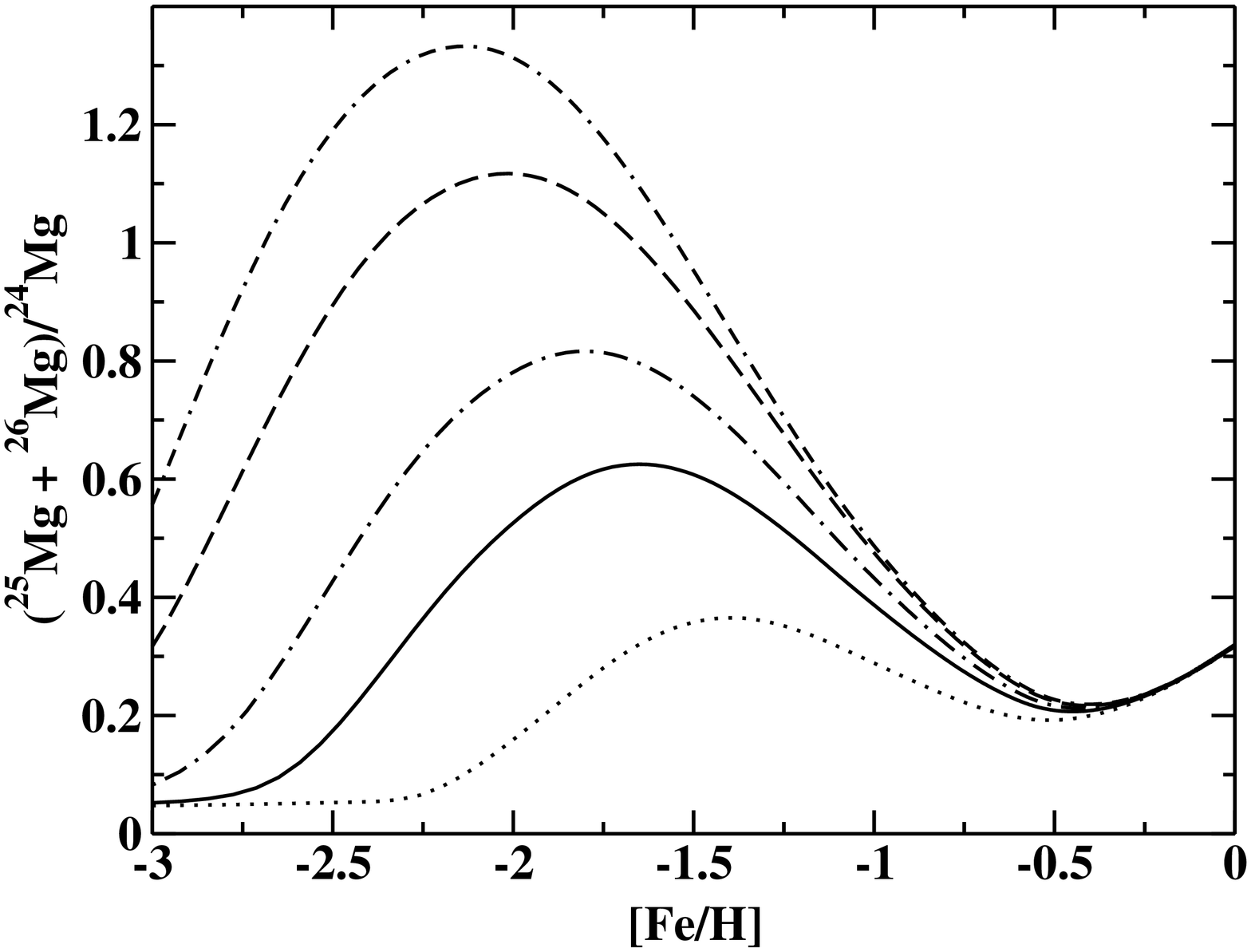,height=9cm,angle=270}
\epsfig{file=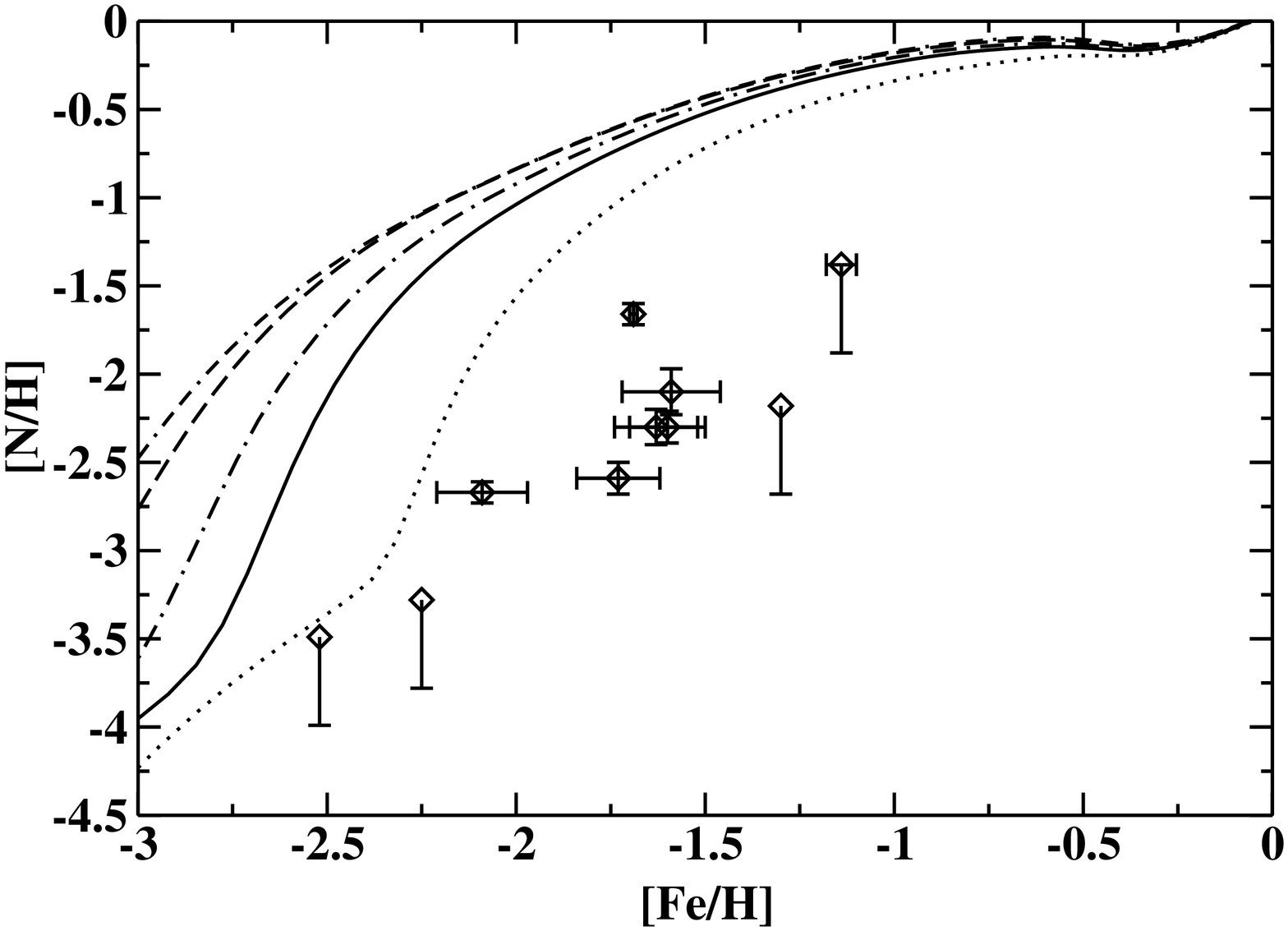,height=9cm,angle=270}}
\caption{Dependence of the magnesium isotopic ratio evolution on dimensionless
width of the log-normal distribution, $\sigma$.  All other
parameters were kept constant except $\sigma$, which had the values
of 0.03 (dotted), 0.07 (solid), 0.1 (dot-dashed), 0.15 (dashed), and 0.2 (dot-dash
-dashed).  
}   
\label{sigmamodel}
\end{center}
\end{figure}

Since  C and N  are products of AGB stars, their observed
abundances in DLA systems  are capable of placing strong 
constraints on this particular chemical evolution model.  In Fig. \ref{fig:n}, we
show the N abundance produced in Model 1 with and without the AGB contribution.  
Comparing the results with observations of the nitrogen
abundance in DLAs indicates that the  AGB enhancement in Model 1 overproduces nitrogen.
Fig \ref{fig:n} also shows that the standard model without an
enhanced IMF (model 0) underproduces nitrogen.  While this 
apparent underproduction may be 
the result of the stochastic nature of the the star formation process, in the context of 
our models at least some enhancement of the IMF for IM
stars seems required by these data.
 The model parameters can in fact be adjusted
to minimize the discrepancy with [N/H], while maintaining the
requisite Mg isotopic enhancement to account for the variation in $\alpha$. 
Indeed, the result shown in Fig. \ref{alfe} is clearly dependent on our choice of the 
six basic model parameters listed in Table 1.
 In order to optimize the model with respect to both its ability to
account for the apparent shift in $\alpha$ and to fit the abundances of the CNO elements,
we vary each of the these parameters. The result of this optimization will be referred 
to as Model 2. The results of these variations are shown in Figs. \ref{stareff} - \ref{sigmamodel},
where we illustrate the influence of each parameter on the shift in $\delta \alpha$
as well as their effect on the nitrogen abundance evolution as a function of [Fe/H]. 
In each of these figures, the solid curve corresponds to Model 1.

In an independent high-resolution observation, the variation of the fine-structure constant was 
also measured by Chand et al.~(2004).  They ascertained that 
$\delta\alpha / \alpha$ = -0.06 $\pm$ 0.06 assuming Solar Mg isotopic abundances. 
In order to completely account for this variation,
a much smaller IMF enhancement would be needed.  A model that 
fits their results also fits the nitrogen
abundance well compared to a model with a standard IMF. 
We have also  performed  an optimized fit
to their independent result.  This is given as  Model 3 in Table 1.

To begin with, the star formation efficiency plays a major role in both
the degree to which the heavy Mg isotopes are produced as well as the degree to which 
nitrogen is produced in these models. 
The parameters for the efficiency of star formation and the infall time 
for the gas can be freely adjusted as well,
 since the galaxies corresponding to the DLAs encompass a wide range of morphologies and conditions.  
Timmes et al.~(1995) determined that the
star-formation efficiency parameter, $\epsilon_{SF}$,
for the Milky Way in the local neighborhood was close to 3.  
Elliptical galaxies are generally regarded as being more efficient in forming stars than spirals, 
while smaller galaxies should be less efficient.  
Although the Mg isotopic ratio is largely insensitive to the star formation efficiency, 
it does affect the position of the peak with respect to [Fe/H].  
A higher star-formation efficiency will produce more massive stars before the 
longer-lived IM stars recycle into the interstellar medium (ISM) as Fig. \ref{stareff} illustrates. 
Relaxing the metallicity requirement for 
Type Ia supernovae would shift this plot and others to higher [Fe/H].  However,
this adjustment can be compensated by reducing the star formation efficiency. 

 In Fig. \ref{stareff}, we show the effect of varying $\epsilon_{SF}$
on  ($^{25}$Mg+$^{26}$Mg)/$^{24}$Mg and [N/H] versus [Fe/H]. 
In addition, we see from Fig. \ref{stareff}b
that indeed the trend in [N/H] abundances is better
reproduced  at low metallicity when $\epsilon_{SF}$ is larger. 

As described earlier, several model parameters affect the size of the IMF enhancement. 
The parameter $A_2$ governs the weighting of the AGB enhancement with respect to the standard Salpeter IMF component 
before the AGB enhancement has been exponentially reduced. As expected, the peak of the Mg isotopic ratio is linearly dependent on the parameter $A_2$ as seen in Fig. \ref{a2}.
 In effect, the Mg isotope peak can be scaled with this parameter. 
 The nitrogen overproduction is tempered when $A_2$ is reduced. 

The time constants, $\tau_1$
 and $\tau_2$ also affect the total number of IM stars beyond the Salpeter IMF alone as shown in Figs. \ref{t1} and \ref{t2}. The constant $\tau_1$ determines the timescale for the onset of the normal component,
 and as one can see, has an important effect on the nitrogen abundance.
At the same time, $\tau_1$  causes the Mg isotope ratio to become less favorable toward
 accounting for the apparent variation in $\alpha$. Fortunately, 
 other parameters can boost the abundance of the heavy Mg isotopes without
 adversely affecting N/H. The time constant, $\tau_2$ which determines how long the IM enhancement lasts is a prime example. By reducing $\tau_1$ (0 corresponds to the standard component
 present at the onset of star formation) and increasing $\tau_2$, we can maintain 
 the high abundance of heavy Mg isotopes and at the same time reduce the nitrogen abundance.

Other parameters do not change the total enhancement of IM stars; rather,
 they affect the distribution. 
The isotopic ratio is strongly dependent upon the center of the log-normal IMF enhancement, $m_c$.  
Because the yield of the Mg isotopes is most efficient 
for progenitor stars with mass between $5 - 6 M_\odot$, 
the Mg peak will rise and fall depending upon the proximity of the peak in the IMF 
to this range. In Fig. \ref{mcmodel1}a, the
model parameters are kept constant except $m_c$.  This figure shows that
independent of the total size of the IM enhancement, the Mg isotopic ratio
can be greatly enhanced by the choice of $m_c$.  The effect of the variation of $m_c$ on the nitrogen abundance is shown in Fig. \ref{mcmodel1}b.

As Fig. \ref{mcmodel1}a illustrates, the Mg ratio requires a value of $m_c$  around 5-6 $M_\odot$, whereas 
models with such high values for $m_c$ 
do more poorly for the evolution of N/H with respect to the data.
To further constrain the
optimum value of $m_c$, we can use a broader distribution of the AGB IMF component. However, the choice of $m_c$ does constrain the allowed range of the
model parameter, $\sigma$, given in Adams \& Laughlin (1996). 
In Fig. \ref{sigmamodel}, we show the corresponding effects of the variation of the width $\sigma$
on the Mg isotopic ratio and nitrogen evolution.

\begin{figure}[ht]
\begin{center}
\mbox{\epsfig{file=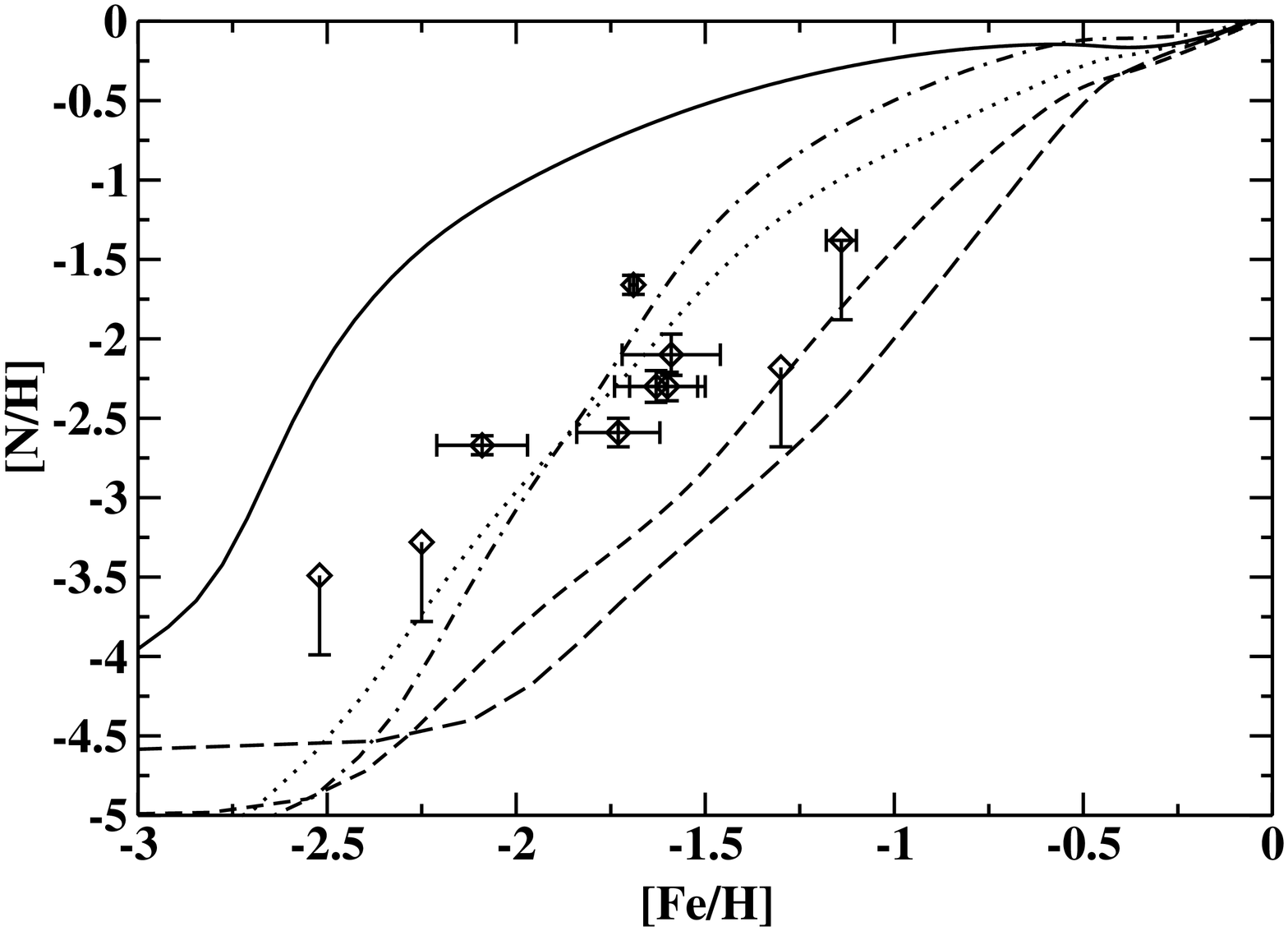,height=10cm,angle=270}}
\caption{ Comparison of nitrogen evolution to observations in DLA systems.  With Model 1 (solid curve), a marginal excess is produced of [N/H]. The standard
IMF corresponding to Model 1 (short dash curve) underproduces N these DLA
 systems.  Models 2 (dot-dash) and 3 (dotted) match the observed [N/H] considerably better than their standard IMF curve (long dash). The data was 
taken from individual abundance ratios of DLA systems (Centurion et al.~2003; 
Pettini et al.~2002; D'Odorico et al.~2002; Molaro et al.~2001; and Levshakov 2001).
}   
\label{nhvsfe}
\end{center}
\end{figure}

Finally, using our understanding of the available parameter space, 
the chemical evolution model was 
adjusted to obtain an optimal concordance with the
measured abundances in DLA systems. While these systems
exhibit a large degree of scatter in their respective abundances, a 
qualitative comparison can be performed.  This comparison may also detail the
consequences of adopting an IMF enhanced with IM stars, especially when
compared to the standard Salpeter IMF model.  Fig. \ref{nhvsfe} shows
a comparison of calculated and observed [N/H] vs. [Fe/H] for the
various chemical evolution models described herein.
The optimized models, 2 \& 3, provide a fair reproduction of the overall trend
in the observed abundances.

\begin{figure}[ht]
\begin{center}
\mbox{\epsfig{file=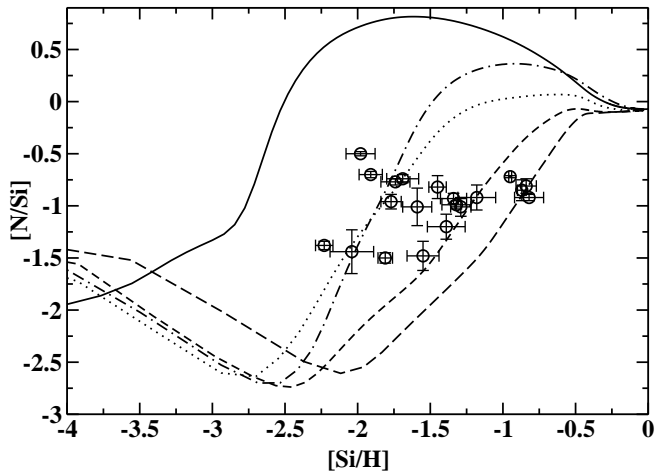,height=10cm,angle=270}}
\caption{ Plot of [N/Si] for [Si/H] for models with and without an AGB enhanced IMF.
 The data is compared to the DLA abundances  of Centurion et al.~(2003).  Model 1 (solid), Model 2 (dot dash), and Model 3 (dotted) are compared to the standard (without an AGB enhanced) IMF models for 1 (short dash) and 2\& 3 (long-dash).
This data points illustrate two plateaus in the distribution of [N/Si] equal
to around -1.5 and -0.9. 
}   
\label{nsivssih}
\end{center}
\end{figure}

Fig. \ref{nsivssih} shows the evolution of nitrogen relative to the alpha element silicon,
which is dominantly produced in Type-II supernovae.  
This figure also illustrates the large abundance dispersions 
typical of DLA systems.  The same models are compared to the [N/$\alpha$] abundance as
[$\alpha$/H] evolves.  The largest sample 
for Si abundances comes from the  20 measurements from Centurion et al.~(2003).  
Although there is a great deal of scatter in these data, it has been interpreted that
these abundances 
show a low dispersion [N/H]=-1.5 plateau and a second high dispersion plateau
for [N/H]=-1.  They contend that primary nitrogen from very massive objects
would not be able to reproduce the lower plateau, while primary-yields from lower-mass stars
do in fact produce this plateau. Both the AGB enhanced chemical evolution
models and the standard IMF model show a plateau, however, the lower plateau
is underestimated by 0.5 dex. This fact may indicate that the adopted WW95
primary yields underestimate nitrogen, or that  an additional source of primary
nitrogen has been overlooked. 
Another important consideration is that 
the Si yield depends sensitively on
the adopted explosion energy and mass cutoff of Type II supernovae.  
The adopted choice of explosion energy in the
no-metallicity model of WW95
leads to a dearth of Si produced. Consequently, this  
leads to  a downward trend at the lowest values of [Si/H] in Fig. \ref{nsivssih}
instead of the observed plateau. 
Centurion et al.~(2003) point out that
the yields of Meynet \& Maeder (2002) match this 
plateau well. They
further contend that the second plateau is due to the addition of secondary 
N from        IM         stars. While we do not observe
a higher plateau in either the enhanced or standard IMF case, the 
enhanced IMF is more consistent with these abundances than the standard IMF. 
While both models suffer from a steep slope in between the two plateaus (the
steep slope can explain the expeditious transition from the lower plateau to the higher one), 
the enhanced IMF model can be easily parametrized to account for 
the range of values of the upper plateau. 
The standard IMF is unable to account
for the broad range of values in the upper plateau.    
However, one must keep in mind that the standard IMF does not
have the parameter degrees of freedom that the enhanced IMF does.

\begin{figure}[ht]
\begin{center}
\mbox{\epsfig{file=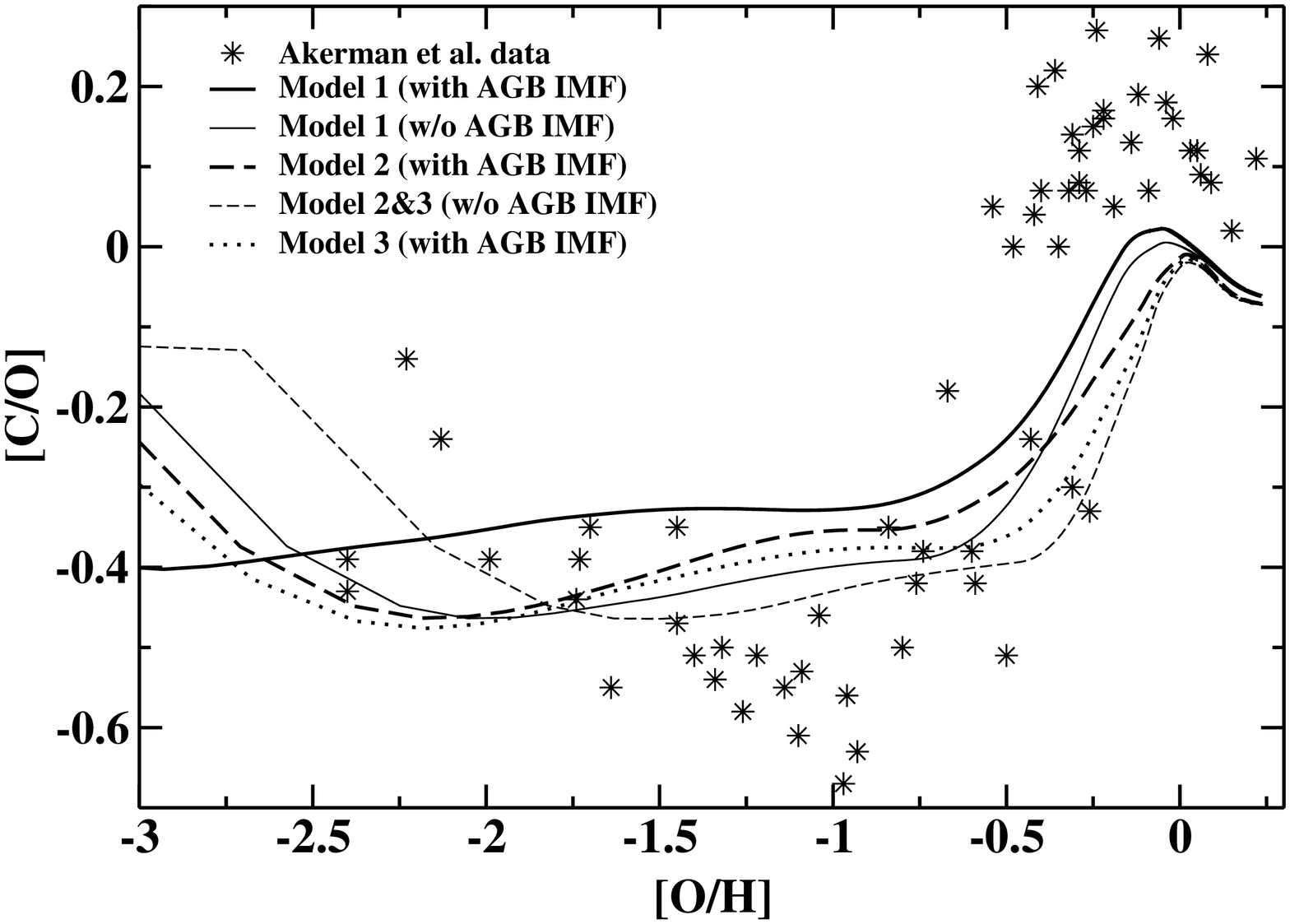,height=10cm,angle=270}}
\caption{ Plot of [C/O] for [O/H] for models with  and without an AGB enhanced IMF.  The data is compared to the stellar abundances (Akerman et al.).  Model 1 (solid), Model 2 (long-dash), and Model 3 (dotted) are compared to the standard (without an AGB enhanced) IMF models for 1 (thin solid) and 2\& 3 (short-dash).
}   
\label{co}
\end{center}
\end{figure}

Since carbon is also a possible product of        IM         stars, we also illustrate the
evolution of C relative to O (primarily from SNII) in Fig \ref{co}.  Here we see that essentially
all of the curves give a similar good fit to the observed abundance (Akerman et al.~2004) of metal-poor halo
stars.  Here we see that all of the models do equally well.  The  basic trend of these data is that
at first the C/O diminishes as supernovae produce more oxygen relative to carbon.  Later
for [Fe/H]$> -0.5$, C/O increases due to the late-time ejection of carbon from low-mass AGB stars.
The enhanced early IMF considered here has little effect on these curves, except for Model 1 in which
substantial early carbon enrichment causes the C/O to be a bit too high near [Fe/H] $\sim -1.5$.
Nevertheless, our optimum models 2 \& 3 do well in comparison to the data.
We note that while it is reassuring that the models fit the data for local halo stars,
chemical models for DLAs and the local neighborhood could be very different.

\begin{figure}[t]
\begin{center}
\mbox{\epsfig{file=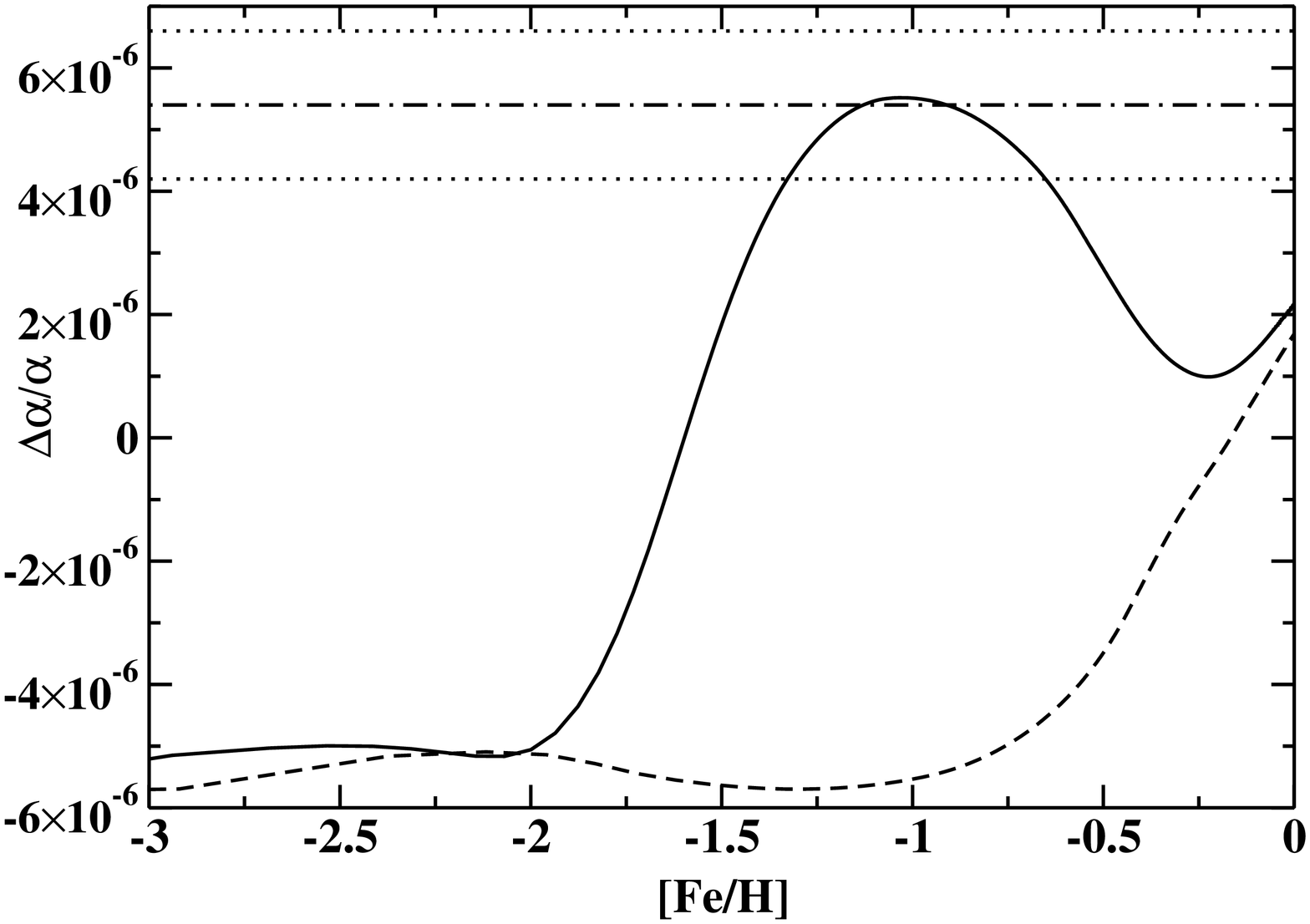,height=9cm,angle=270}
\epsfig{file=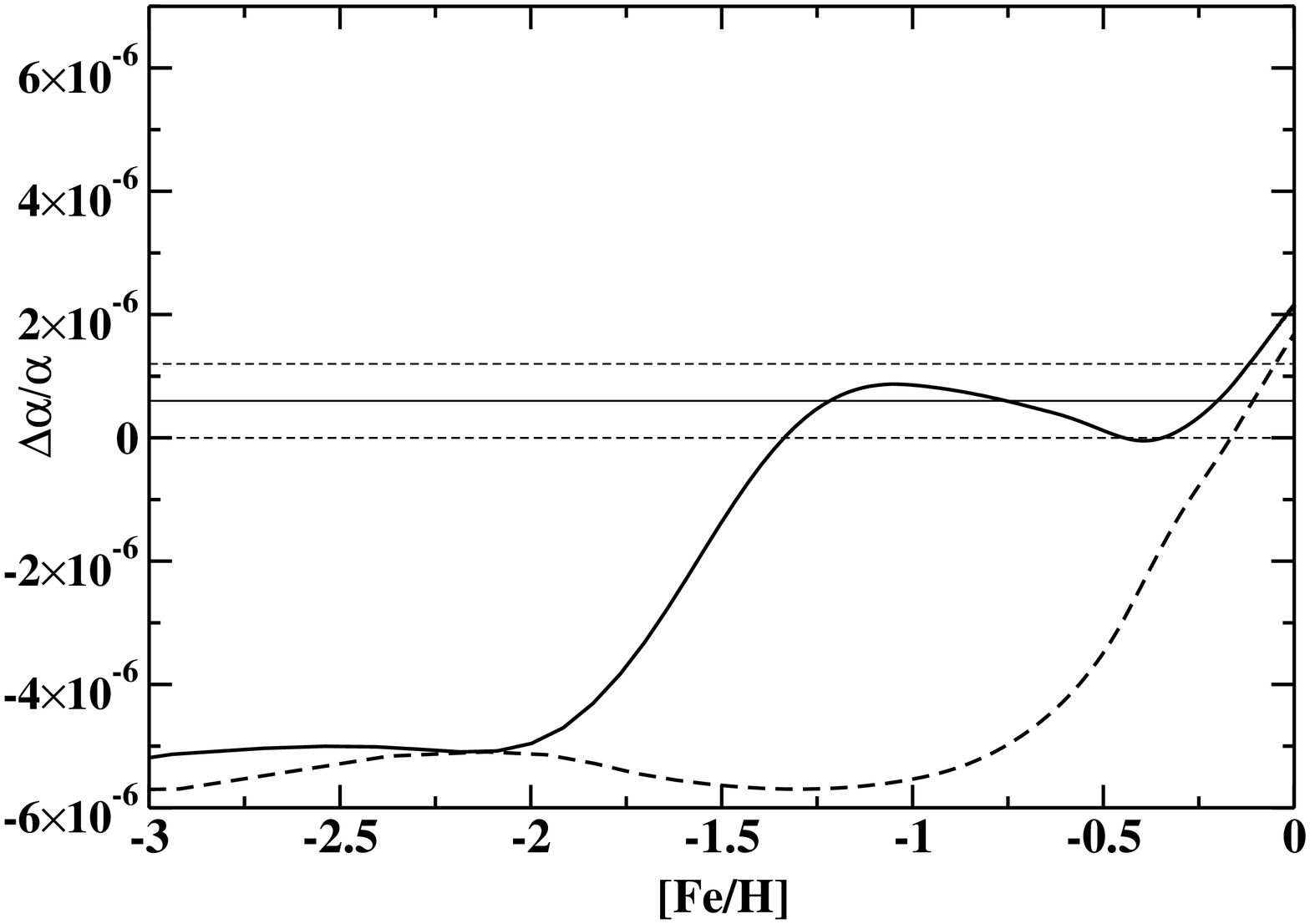,height=9cm,angle=270}}
\caption{As in Fig. \protect\ref{alfe}, the shift in the fine-structure constant 
due to isotopic abundance variations in Mg corresponding to Model 2 (left), the  fit to the 
Murphy et al.~(2003a) result and to Model 3 (right), the  fit to Chand et al.~(2004).  
}   
\label{alfeopt}
\end{center}
\end{figure}

Finally, in Figs. \ref{alfeopt}a and b, we summarize our estimated shift in deduced $\delta \alpha/\alpha$
relative to that obtained from a Solar Mg isotopic composition.  It is clear that either of these
models provide a possible alternative to the interpretation of a time varying fine-structure constant.
Furthermore, the properties of these models of early AGB enhancements to the IMF
are suggested  
by various abundance measurements in DLA systems and metal poor stars. In this interpretation,
the MM method results have provided important new evidence into the star formation that occurred in the
early epochs of DLAs.

Before concluding, we make two final points.  First, we emphasize that there is 
considerable scatter in the data. Scatter is found in both N/H data as well as
the data from which Murphy et al.~(2003a) and Chand et al.~(2004) infer
a value for $\delta \alpha / \alpha$ for each individual absorber.
Our model can only be viewed as an average star formation history
over many individual DLAs.  Indeed, because of the stochastic nature
of star formation, particularly at early times, we might expect 
large variations in the production of IM stars, and hence the
production of the heavy Mg isotopes.  For this reason, we 
believe that the MM method may provide a unique
window to the star formation history in DLAs.  
Second, we also emphasize that, although the Chand et al.~(2004) data
are consistent with no variation in $\alpha$, this is only so, because 
they chose Solar abundances for $^{25,26}$Mg.  As one can
see from Fig. \ref{alfeopt}b, Model 3 which fits this data still requires
a strong IM component in order to produce nearly Solar isotopic abundances of Mg.
 
\section{Summary and Conclusions}

We have made a study of possible relations among stellar nucleosynthesis,
the galactic chemical evolution of damped Lyman-$\alpha$ systems, and  
the apparent detections of a time-varying fine structure constant.
In particular, we have explored the important effects of high temperature
thermonuclear burning in low-metallicity AGB stars.  We have
shown that ejecta from these stars could have had a dominant effect 
on the early galactic chemical evolution of the crucial 
Mg isotopes in DLA systems.

We have explored a variety of models in which the early initial mass function
favors the formation of        IM         stars.  Such an enhanced
contribution from early IM stars allows for sufficient modification
of the Mg isotope ratios to explain the many-multiplet results
without a time-varying fine structure constant.  Such a modified IMF may 
to some extent be a simple parametrization of uncertainties in 
theoretical estimates of ejected yields from AGB stars, but it is 
motivated by both theoretical and observational constraints.

  To compare with the MM method results, we have utilized 
an approximate treatment that qualitatively relates computed Mg
isotopic abundances to the deduced fine structure constant 
for DLA systems in the redshift range
of $0.5 < z < 1.8$.  Although this is only an approximation to the MM
approach, we have shown that it reproduces the  basic conclusions
of detailed analysis (Murphy et al.~2003a). 
There is a real need to redo the
MM method analysis in the context of evolving
isotopic abundances as derived here.  Incorporating isotopic
variation would help
to better quantify the need (or lack thereof) for IM 
stars and AGB nucleosynthesis in DLA systems. We hope
that the present work will stimulate efforts along this line.
 
 In the context or our schematic analysis, 
 we have explored a variety of chemical evolution models with an eye toward 
unraveling the time-varying alpha mystery while still satisfying the
available constraints from observed elemental abundances in DLA systems.
We have concentrated on the chemical evolution of N abundances,
which are also produced in AGB stars. We also considered
C as well as  O and Si largely from Type II supernovae, 
and Fe from SNIa and SNII. We find that the observed 
high nitrogen abundances in DLA systems indeed  confirm
 the need for enhanced ejecta from low-metallicity AGB stars.
Even so, our previous model (Ashenfelter et al.~2004), which attempted to
explain the MM results of (Murphy et al 2003a)
 tends to overproduce nitrogen and is therefore 
constrained by the observations.  In this paper,
however, we report on a parameter search which considers 
both data sets.  We find  a new optimum
model (Model  2 in the present work) which simultaneously fits 
the observed N/H, C/O, and N/Si trends vs [Fe/H] while still 
eliminating the need for the time-varying fine structure constant
as deduced from the Murphy et al.~(2003a) data for systems 
in the redshift range $0.5 < z < 1.8$.

At the same time, we have also constructed a new model (Model 3)
that can account for the results of the  independent 
MM method analysis of Chand et al.~(2004), 
which indicate smaller apparent variations 
in the fine structure constant.  Even though these authors
claim results that are consistent with no variation in $\alpha$, this conclusion 
is based on the assumption that the Mg isotopic ratio is equal to the 
Solar one.  As we have shown, in order to obtain a Solar isotope 
ratio at low metallicity, we must again rely on the role of IM stars and AGB nucleosynthesis.

 One conclusion of the present study is that important tests can
be made of the hypothesis that AGB nucleosynthesis can account for the
apparent variation in the fine structure constant.  The best measurement
(though probably impossible) would be to directly detect Mg isotopic abundances 
from spectral lines.  In our picture, the apparent variations in $\alpha$ should
correlate directly with the fraction of heavy Mg isotopes. Conversely,
if heavy magnesium abundances are significantly 
depleted relative to Solar, then the
MM method results are actually understating the variation
in the fine structure constant. 
Furthermore, if sufficiently precise data could be obtained to distinguish the
$^{25}$Mg and $^{26}$Mg abundances, then large enhancements 
of $^{26}$Mg observed in some systems could be
indicative of Mg production that is specifically attributable to the Mg-Al
cycle.  If so, large enhancements of $^{26}$Mg may also be anti-correlated with Al abundances.

We further suggest that nitrogen (and/or carbon) abundances provide an easier test of the
present hypothesis with regards to an enhanced IMF for 
low-metallicity        IM         stars.
Nitrogen  should be measured and correlated with $\delta \alpha/\alpha$ 
in the same DLA systems to which the MM method is applied. A correlation
of [N/H] with the largest variations in $\alpha$ would argue in favor
of the present hypothesis.
 As another test of an enhanced IMF,
 the highest [Fe/H] or [Si/H] (Si and Fe are very
correlated) in DLA systems should exhibit a significant over abundance of 
[N/$\alpha$] if our IMF is correct. 
One caveat in using Si abundances is
its dependence on the explosion conditions for the Type II supernovae. 
Future work regarding these tests should examine the consequences of adopting
these different explosion criteria and compare with different yield models.

Determining whether or not a metallicity condition exists for Type Ia 
supernovae is decisively related to an IMF enhanced with intermediate-mass
stars. It may very well be that the reduced efficiency of Type Ia supernovae
is offset
by the enhanced numbers of intermediate mass stars, which may satisfy the 
previous work of Matteucci \& Recchi ~(2001) and the metallicity conditions of
Nomoto et. al ~(2003).  Comparison of abundances of N, C, and Mg with [O/H] 
may be able to test the viability of IM mass enhanced IMF models independent
of Sn Ia rates. 

One should interpret the 
large scatter in the inferred variation in $\alpha$, as well as the observed variation in
element abundances like Mg, in the context of the stochastic nature of the 
star formation process at low metallicity.  Our models simply
represent a global average, and the contribution of IM stars
may very well vary in individual DLAs.

Clearly, more work needs to be done in constraining the effect of
chemical evolution on the interpretations of DLA observations. 
Nevertheless, we have established that at least
some fraction of the deviations in the fine-structure
constant deduced from the MM method 
could be due to chemical evolution effects.
Among work that needs to be done, more
stellar models of low-metallicity AGB stars  over a broader mass 
and metallicity range are needed to quantify the model predictions,
particularly in the extrapolated mass range of $7 <$ M $< 12$ M$_\odot$
 between the AGB and Type II regimes. In particular, it would be very useful
 to have have a full complement of stellar yields (including C,  N and Mg)
 in the IM range derived from a self-consistent set of stellar models. 
Out of necessity, we supplemented the Mg yields of Karakas \& Lattanzio (2003)
with the Padova CNO yield models; however, a self-consistent model would
more accurately quantify the correlation of Mg from AGB sources to N.

 At the same time, more observations of abundances in DLA systems 
are needed. Most importantly, the DLAs that are used in the MM-method should
 have their associated abundances quantified in order to see if there
 are correlations between the apparent variations in $\alpha$ and various abundances.
Additionally, more atomic physics work needs to be done to 
more accurately quantify the possible isotopic shifts in the absorption lines 
(particularly for the high-$z$ data).

The MM method has presented a very important question 
as to whether the
fine structure constant varies with time. 
It is hoped that the present study will stimulate further 
efforts along all of the above  lines with a goal of
clarifying  this important physical question.
If the fine structure constant does vary in time or space,  it
provides an important window into the physics beyond the standard model.
As such, the chemical evolution effects described herein should be
carefully quantified to reduce the systematic uncertainties in the deduced
result.  Even if our
chemical-evolution interpretation of the MM results prove to
be verified, the MM method will have provided 
valuable insight into the mysteries of early cosmic 
 star formation and galaxy  evolution.


We thank V. Flambaum for helpful conversations.
The work of K.A.O. was partially supported by DOE grant
DE--FG02--94ER--40823.
 Work at the University of Notre Dame was supported by the
U.S.~Department of Energy under Nuclear Theory Grant DE-FG02-95-ER40934. 
One of the authors (T.A.) also  wishes to acknowledge
partial support from NSF grant PHY02-16783 through the
Joint Institute for Nuclear Astrophysics physics (JINA).

\end{document}